\documentclass[sigconf]{acmart}
\renewcommand\footnotetextcopyrightpermission[1]{}
\setcopyright{none}
\PassOptionsToPackage{table}{xcolor}

\usepackage{float}
\usepackage{booktabs}
\usepackage{adjustbox}
\usepackage{multirow}
\usepackage{tikz}
\usepackage{colortbl}
\usepackage{array}
\usepackage{makecell}
\usepackage{siunitx}
\usepackage{subcaption}
\usepackage{xparse}
\usepackage{listings}
\usepackage{tabularx}
\usepackage{longtable}
\usepackage{verbatim}
\usepackage{pdflscape}
\usepackage{xfp}
\newcommand{\fivestars}{$\bigstar$ Ratings}

\lstdefinestyle{prompt}{
  basicstyle=\ttfamily\scriptsize,
  columns=fullflexible,
  keepspaces=true,
  breaklines=true,
  breakatwhitespace=true,
  frame=single,
  rulecolor=\color{black!20},
  showstringspaces=false
}
\definecolor{PosCol}{RGB}{240,180,100}   % red
\definecolor{ZeroCol}{RGB}{255,255,255} % white
\definecolor{NegCol}{RGB}{130,210,220}   % blue
\definecolor{tealHigh}{RGB}{240,180,100} % medium-light teal
\definecolor{rowgray}{gray}{0.94}

\newcolumntype{C}[1]{>{\centering\arraybackslash}m{#1}}
\newcolumntype{L}[1]{>{\raggedright\arraybackslash}m{#1}}

\usepackage{xfp}
\newcommand{\WCell}[2]{%
  \begingroup
  \edef\rv{#2}%
  \edef\pct{\fpeval{min(100, round(200*abs(\rv),0))}}%

  \ifdim \rv pt < 0pt
    \edef\WCellColorSpec{NegCol!\pct!ZeroCol}%
  \else
    \edef\WCellColorSpec{PosCol!\pct!ZeroCol}%
  \fi
  \expandafter\cellcolor\expandafter{\WCellColorSpec}%
  \strut #1%
  \endgroup
}

\newcommand{\Pos}[2]{\cellcolor{PosCol!#1!ZeroCol}\strut #2}
\newcommand{\Neg}[2]{\cellcolor{NegCol!#1!ZeroCol}\strut #2}

\newif\ifshowsuggestions
\showsuggestionstrue % set to \showsuggestionstrue to bring all \suggest{} comments back

\AtBeginDocument{%
  }

\begin{document}

\title{Beyond Satisfaction: Learning Associations Between Content, Reviews, and Well-Being}

\author{Aaron Marker}
\email{aaron.marker@vanderbilt.edu}
\affiliation{%
  \institution{Vanderbilt University}
  \country{USA}
}
\author{Joel Lehman}
\affiliation{%
  \institution{University of Oxford and Cosmos Institute}
  \country{USA}
}

\author{Oscar Kjell}
\affiliation{%
  \institution{Lund University}
  \country{Sweden}
}

\author{H. Andrew Schwartz}
\affiliation{%
  \institution{Vanderbilt University}
  \country{USA}
}

\renewcommand{\shortauthors}{Marker et al.}

\begin{abstract}
Recommendation algorithms shape what people consume across the web, yet they are primarily optimized for star ratings, likes, and sentiment, signals of user \textit{satisfaction}. Whether these signals capture what is good for users, beyond momentary preference, remains largely unexamined.
\textit{Well-being} offers an alternative target grounded in psychology that extends beyond satisfaction.
We operationalize it as immediate and enduring expressions of the five PERMA facets in reviews. 
Empirically, immediate positive emotion and engagement cluster as more \textit{hedonic} (pleasure-oriented), while relationships, meaning, accomplishment, and enduring positive emotion and engagement behave as more \textit{eudaimonic} (meaning- and growth-oriented).
Analyzing 13.7 million Goodreads reviews and the content of 16 thousand books, we study the extent to which both immediate and enduring expressions of PERMA in the reviews are distinct from traditional satisfaction targets and which features of book content are associated with them.
First, satisfaction signals (star ratings and review sentiment) correlate moderately with \textit{immediate} hedonic well-being ($r = .43$--$.50$) but less with its \textit{enduring} expressions ($r \leq .28$), and have almost no association with eudaimonic facets ($r \leq .09$).
Second, book content features are more strongly associated with PERMA facets ($r$ up to $.46$) than with satisfaction signals ($|r| \leq .15$).
Content associated with immediate hedonic well-being differs from that associated with enduring and eudaimonic well-being: third-person pronouns (e.g., `he', `she') and past-focused language relate positively to immediate engagement ($d=1.14$ and $1.05$) but negatively to every enduring facet ($d$ down to $-1.58$ and $-1.32$), while prosocial (e.g., `help', `care') and achievement-oriented (e.g., `work', `better') language relate to relationships and accomplishment ($d=.45$ and $.96$).
These findings suggest that satisfaction signals capture only part of the well-being users express, and that well-being-informed targets could lead platforms to recommend substantially different content.

\end{abstract}




\maketitle

\section{Introduction}
Because recommendation algorithms shape what billions of people consume on the Web, whether their optimization targets serve users' long-term well-being is a central question for a responsible Web.
Web platforms often optimize for user satisfaction using signals such as ratings, likes, and sentiment \cite{zhang2014explicit,zhao2019recommending,kang2023llms,gao2022kuairand, zhang2019deep}. Psychological research characterizes an individual's well-being as a multidimensional construct that extends beyond short-term positive affect or satisfaction to include facets such as meaning, relationships, and accomplishment \cite{seligman2011flourish, diener2018advances}. Beyond its intrinsic value, well-being predicts broader life outcomes, including physical health, longevity, social relationships, and work performance \cite{diener2018advances}.
Whether the satisfaction signals that platforms rely on capture these broader well-being dimensions remains largely unexamined.

\begin{figure}[t!]
     \centering
     \includegraphics[width=.48\textwidth]{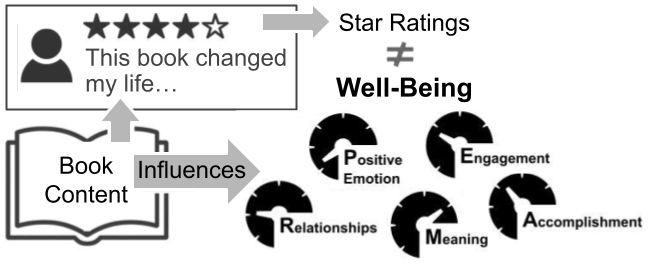}
     \caption{Expressed well-being is modeled as a latent construct inferred from multiple complementary facets, analogous to an aircraft dashboard that requires integrating multiple indicators rather than relying on any single measure. Star ratings and expressions of well-being capture distinct signals in book reviews, each associated with different concepts in the reviewed book. }
     \label{fig:SpiritFigure}
\end{figure}

Language-based assessments, which use machine learning to estimate psychological constructs from text, offer a scalable way to measure well-being beyond coarse signals such as ratings or sentiment \cite{kjell2019semantic, kjell2022natural, schwartz2016predicting}. We apply such assessments to book reviews, then use differential language analysis (DLA) \cite{schwartz2013personality} to study how the content a user consumes is associated with the well-being expressed in their review (\textit{expressed well-being} in this work refers to improvement in a PERMA facet that a reviewer attributes to the book). We focus on book consumption as a tractable domain of non-interactive content engagement: readers engage deeply with fixed texts and often produce long-form reviews that capture nuanced reflections on their experience. Using a large corpus of Goodreads reviews paired with book text from the HathiTrust Digital Library, we analyze how readers describe the well-being benefits they attribute to engaging with a single work (Figure~\ref{fig:SpiritFigure}). For this domain specifically, prior research also suggests that reading, particularly fiction, can influence mental health and well-being \cite{arslan2022story, carney2022fiction}, and a growing area of work has conducted large-scale analysis of book text \cite{hobson-etal-2024-story, troiano-vossen-2024-clause}. Reviews provide reflective, user-centered language focused on individual experience, allowing us to link properties of content to expressed well-being with fewer conversational confounds than interactive platforms.

We operationalize well-being using the PERMA framework \cite{seligman2011flourish} and estimate facet scores across 13.7 million Goodreads reviews using both LLM and human annotation for supervised modeling. We then relate book-level well-being, aggregated across reviews, to linguistic features of full book texts.
We address two research questions: \textbf{(RQ1)} Do the ratings and sentiment that platforms optimize capture eudaimonic, as well as hedonic, well-being expressed in reviews? \textbf{(RQ2)} What book content is associated with eudaimonic versus hedonic expressions of well-being?
Rather than proposing a new recommender system, we focus on the measurement question that precedes one: whether the signals such systems rely on capture the well-being users express, what they miss, and whether targeting well-being would surface different content.
Our contributions are threefold.
First, we show that ratings and sentiment only loosely correlate with most facets of well-being and are more closely aligned with immediate and hedonic expressions of well-being than with enduring and eudaimonic ones, and we identify book concepts associated with higher and lower well-being.
Second, we develop a large-scale framework for analyzing well-being in reader reviews by integrating psychological theory with language-based assessment of reviews and differential language analysis of book content.
Finally, we provide a new dataset containing 3,500 human-annotations and 50,000 GPT-5-annotations for expressed well-being in book reviews.

\section{Related Work}
\label{sec:related-work}
\paragraph{Satisfaction Signals on Web Platforms} Star ratings, a common signal of user satisfaction (here, satisfaction being a user's evaluation of a specific item, not life satisfaction as a component of well-being~\cite{diener2018advances}), have been a central recommendation target for more than a decade \cite{koren2009matrix} and remain one today \cite{kang2023llms}. Ratings also shape choices directly: on platforms such as Yelp and Amazon, displayed average ratings measurably affect sales \cite{chevalier2006effect, luca2016reviews}. Beyond ratings, platforms such as YouTube optimize engagement signals including watch time and likes \cite{covington2016deep, zhao2019recommending, gao2022kuairand}. A growing body of work draws on review language, from sentiment analysis \cite{zhang2014explicit} to language-model-based recommendation \cite{geng2022recommendation, hou2026bridging}, but uses it to predict preference rather than to assess an item's effect on the user. Even work targeting a user's long-term welfare typically operationalizes it only as sustained engagement or return visits \cite{wu2017returning, zou2019reinforcement, zhao2025enhancing}. Each of these signals captures how much a user liked or engaged with an item, but none is designed to capture the broader dimensions of well-being.

\paragraph{Beyond Satisfaction: Well-Being in Recommendation}
Engagement and satisfaction signals may not reflect what is good for users \cite{milli2021optimizing}: when in-the-moment choices diverge from what users actually want, more engagement can leave them less happy \cite{kleinberg2024challenge}. Researchers have therefore called for recommenders that support users' ``better selves'' \cite{ekstrand2016behaviorism, knijnenburg2016recommender} and treat well-being as an explicit target \cite{stray2020aligning}, yet measuring well-being in production remains an open problem \cite{stray2024building}. Field experiments measure it directly, reranking participants' feeds and surveying them over weeks or months \cite{piccardi2025reranking, stray2026prosocial}; such designs provide causal, user-level evidence but cannot attribute effects to individual items. Work on hedonic and eudaimonic media has shown that these qualities can be predicted for individual items and users, but captures two dimensions from small questionnaire samples \cite{motamedi2022prediction, motamedi2024predicting}. We complement these approaches by measuring well-being from what users themselves write about individual items, across five theory-grounded facets and their immediate and enduring expressions. Validated against human annotation, our models scale this measurement to 13.7 million reviews, enabling item-level analysis that neither surveys nor field experiments can provide.

\paragraph{Media Consumption and Well-Being}
Prior research across media domains suggests that the psychological impact of content depends not only on the quantity of exposure but also on the nature of consumed content. Studies of social networking site (SNS) use report associations with depression, anxiety, and lower well-being \cite{keles2020systematic, vidal2020social, agyapong2025effects, shannon2022problematic, nagata2025social}, though effects depend on how users engage with content rather than usage volume alone \cite{MARCIANO2024100331, verduyn2022social, verduyn2017social}. In particular, passive consumption—observing content without direct interaction—seems to have the most potential for harm and has been linked to more negative mental health outcomes ~\cite{tiggemann2015exercise, fox2016selective}. Importantly, the influence of consumed content on well-being extends beyond social media. News consumption, for instance, is increasingly associated with negative emotional outcomes, frequently leading citizens to avoid it \cite{de2020news,newman2025digital}. Similarly, research has shown that reading books (the domain studied here) can also influence mental health and well-being \cite{arslan2022story, carney2022fiction}. This work suggests that what people consume, not just how much, shapes well-being, motivating our analysis of which content features are associated with well-being expressed in reviews.

\paragraph{Hedonia vs Eudaimonia}
Hedonia is defined broadly as experiencing pleasure and avoiding pain~\cite{ryan2001happiness}. In contrast, Aristotle described eudaimonia as activity in accordance with virtue; psychological accounts operationalize it as meaning, growth, and self-realization~\cite{ryff1989happiness, waterman1993two, huta2014eudaimonia}, emphasizing psychological functioning over pleasure~\cite{ryan2001happiness}. 
Marciano et al., for example, examine meaning and self-realization in social media use, highlighting how hedonic pleasure can be disconnected from eudaimonia~\cite{MARCIANO2024100331}. Yet whether the two are empirically distinct is debated, as hedonic and eudaimonic well-being overlap substantially in self-report data~\cite{kashdan2008reconsidering, disabato2016different}. If satisfaction signals primarily reflect pleasure, they may capture hedonic well-being while missing eudaimonic facets such as meaning and accomplishment. RQ1 tests both whether these facets separate in reviews and whether satisfaction signals track each.

\paragraph{Immediate vs Enduring}
A second distinction concerns how long well-being changes last. Hedonic adaptation describes the tendency to return to baseline following positive or negative life events~\cite{frederick1999hedonic}, although adaptation is often incomplete~\cite{diener2006beyond}. More durable improvements are associated with intentional activities such as cultivating relationships and pursuing intrinsic goals~\cite{lyubomirsky2005pursuing, sheldon2012challenge}. Similarly, Baumeister et al. find that happiness is tied to the present, whereas meaning involves integrating past, present, and future~\cite{baumeister2016some}. Ratings and sentiment capture how much a reader liked a book, but not whether its effects were transient or lasting. We therefore annotate both immediate and enduring expressions of well-being, distinguishing transient reactions from indications of more durable change.

\paragraph{PERMA}
To characterize well-being expressed in reviews more specifically than a single overall score, we use a small set of broad facets that can be annotated and modeled from text. We adopt Seligman's PERMA theory, which defines five facets: Positive Emotion, Engagement, Relationships, Meaning, and Accomplishment~\cite{seligman2011flourish}. PERMA overlaps with other well-being frameworks, spans hedonic and eudaimonic dimensions, and has been modeled from language~\cite{schwartz2016predicting}. Positive Emotion is primarily hedonic and Relationships, Meaning, and Accomplishment primarily eudaimonic, while Engagement's placement is debated~\cite{heshmati2022assessing}. Although PERMA facets overlap substantially in self-report~\cite{goodman2018measuring}, whether they separate in naturally occurring language remains open; we therefore examine empirically how the facets group in reviews.

\paragraph{Measuring Psychological Constructs from Language}
Language offers a scalable window into psychological states. Supervised language-based assessments train models to estimate psychological constructs from text and can converge closely with established rating scales \cite{kjell2019semantic, kjell2022natural}; they have been used to estimate PERMA facets of individual well-being \cite{schwartz2016predicting} and the subjective well-being of communities \cite{jaidka2020estimating}. Differential language analysis (DLA) instead identifies the words and topics that distinguish psychological outcomes \cite{schwartz2013personality, kern2014online}. On web platforms, however, review language has been used mainly to predict preference rather than to estimate psychological outcomes for the user. We apply language-based assessment to reviews to estimate expressed well-being at platform scale, and DLA to identify the book content associated with it.

\section{Datasets}
\label{sec:datasets}

\paragraph{Reviews}
We use the UCSD Goodreads dataset \cite{wan2018item}. After filtering for English (>45\% English words, using langid) and minimum length (>50 characters), it contains 13.7 million reviews, which we use for all review-level analyses.

\paragraph{Book Content}
To relate reviews to the content of the books themselves, we use the HathiTrust Research Center Extracted Features dataset (EF 2.5), which provides page-level word counts for both copyrighted and public-domain books for research use. Of the 73,168 Goodreads books matched to HathiTrust, we keep those whose text is verified as English by per-page language detection (at least 25\% of pages; 65,943 books). Because Goodreads lists editions of the same work separately, we pool reviews across editions of each work, excluding non-English editions so that reviews of translations are not attributed to the English text. Retaining works with more than five pooled reviews yields 16,341 works and 1,488,774 reviews for our book-content analyses. Genre distributions for these works are in Appendix~\ref{sec:genreTables}.

\section{Methods}
\label{sec:methods}

\subsection{Annotation}
\begin{figure}[t!]
     \centering
     \includegraphics[width=.48\textwidth]{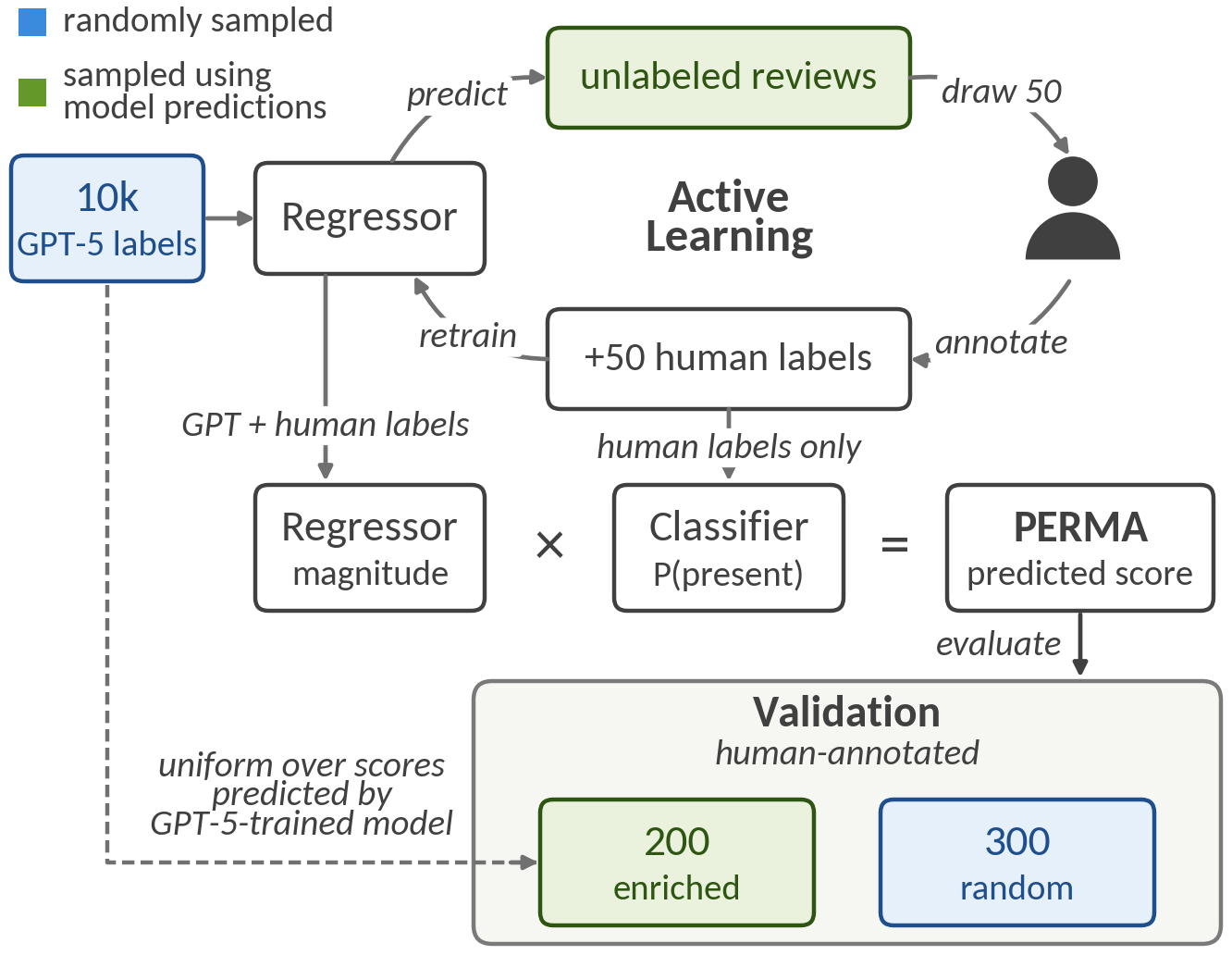}
     \caption{Training and validation pipeline. A regressor trained on GPT-5 labels selects reviews for human annotation over four (active learning) rounds of 50 reviews, retraining after each round. The final regressor uses GPT-5 and human labels; a separate presence classifier uses human labels only, and the PERMA score is their product. Both validation sets are human-annotated and disjoint from training: 300 randomly sampled reviews, and an enriched set of 200 per facet drawn uniformly across predicted score bins from a separate pool of 50,000 reviews.}
     \label{fig:AnnotationMethod}
\end{figure}

We annotate reviews for the five PERMA facets (Section~\ref{sec:related-work}), distinguishing immediate from enduring expressions of each: a reviewer may say they were engrossed in a book without expressing any lasting change in engagement as a result. Because each review is written at a single point in time, enduring scores capture the reviewer's own account of lasting change attributed to the book, not change observed over time. Because manual annotation of 13.7 million reviews is impractical, we follow prior work on prompt-based LLM annotation of rare constructs~\cite{mangalik-etal-2025-capturing} and use GPT-5 (\texttt{gpt-5} via OpenAI's Batch API, default parameters) to annotate 10,000 reviews per facet (50,000 total). Figure~\ref{fig:AnnotationMethod} shows the full annotation and training pipeline.

\begin{table}[htbp]
\small
\centering
\begin{adjustbox}{max width=.6\textwidth}
\setlength{\tabcolsep}{6pt}
\begin{tabular}{llcc}
\toprule
\multicolumn{4}{c}{\textbf{Model Performance}} \\
\midrule
\multicolumn{2}{l}{\textbf{Language task}}
& \makecell{\textbf{Deattenuated} \\ \textbf{Pearson $r$}}
& \textbf{AUC}\\
\midrule
\multirow{2}{*}{\textbf{Pos. emo.}}
 & Enduring  & \textbf{.817} & \textbf{.879} \\
 & Immediate & \textbf{.724} & \textbf{.965} \\
\midrule
\multirow{2}{*}{\textbf{Engage.}}
 & Enduring  & \textbf{.540} & \textbf{.844} \\
 & Immediate & \textbf{.630} & \textbf{.802} \\
\midrule
\multirow{2}{*}{\textbf{Relation.}}
 & Enduring  & \textbf{.870} & \textbf{.925} \\
 & Immediate & \textbf{.831} & \textbf{.898} \\
\midrule
\multirow{2}{*}{\textbf{Meaning}}
 & Enduring  & \textbf{.417} & \textbf{.799} \\
 & Immediate & \textbf{.512} & \textbf{.849} \\
\midrule
\multirow{2}{*}{\textbf{Accomp.}}
 & Enduring  & \textbf{.743} & \textbf{.883} \\
 & Immediate & \textbf{.758} & \textbf{.880} \\
\bottomrule
\end{tabular}
\end{adjustbox}
\caption{Model performance across PERMA facets on the enriched test sets. Deattenuated Pearson $r$ corrects the correlation for measurement unreliability via each facet/outcome's ICC(2,2) (Section~\ref{sec:validation}): $r_{\text{true}} = r / \sqrt{\text{ICC}(2,2)}$. AUC is zero-vs-nonzero classification performance on the same human annotated labels. Performance is moderate to high except for Meaning, the facet with the lowest inter-annotator agreement. All values are statistically significant (Benjamini-Hochberg FDR correction, $q<.05$~\cite{benjamini1995controlling}). Uncorrected correlations, raw accuracy, random-sample performance, and ICC values are in Appendix~\ref{sec:full-model-performance}.}
\label{tab:model-performance}
\end{table}

The prompt draws on descriptions from the PERMA Profiler and \textit{Flourish} \cite{butler2016perma, seligman2011flourish}. For each facet, it asks for an Immediate score (``How much immediate or near-term improvement (within the period of reading or reflecting on the book) is evident for this facet?'') and an Enduring score (``How much deep, long-term, or life-changing improvement in this facet (a fundamental shift in worldview, habits, or identity) is there for the reviewer as a result of reading the book?''). Based on pilot testing, we refined the definitions to separate commonly confused cases: liking a book alone counts as Positive Emotion but not Engagement, interest in continuing a series counts only minimally toward enduring Engagement, and Meaning requires that the book affected the reviewer's own sense of meaning. The full prompt is in Appendix~\ref{sec:model-prompt}.

\subsection{Human Annotation}
Two annotators with backgrounds in NLP and psychology scored reviews for immediate and enduring expressions of each facet; their scores were averaged. Because nonzero examples are rare, random sampling yields few informative labels, so we used active learning over four rounds of 50 reviews per facet. In each round, a regressor trained on the GPT-5 labels selected reviews predicted to score high (first two rounds) or near the zero/nonzero boundary (last two rounds), and was then retrained with the new human labels, yielding 200 human-labeled training reviews per facet (Figure~\ref{fig:AnnotationMethod}). All test reviews, including the pool from which enriched test sets are drawn (Section~\ref{sec:validation}), are disjoint from all training data.

\subsection{Scaling to the Full Corpus}
To label the remaining reviews, we train language-based assessment models on the 10,000 GPT-5 labels per facet and the human-labeled training pool. We embed all reviews using an open-source sentence transformer, mixedbread's large embedding model~\cite{emb2024mxbai, li2023angle}, selected for its high performance on the MTEB benchmark relative to its parameter count and for the strong performance of document-tuned models on psychological prediction tasks~\cite{marker2026evaluating}.

Well-being facets are heavily zero-inflated (most reviews show no evidence of a given facet), so a single regressor fit directly to GPT-5 scores wastes most of its capacity distinguishing among degrees of zero. We instead fit a \emph{two-part model} per facet and outcome (immediate/enduring), mirroring standard approaches to zero-inflated outcomes: (1) a \emph{presence classifier}, a logistic regression on a 20-component PCA projection of the embeddings, trained only on human-labeled zero-vs-nonzero judgments to predict $P(\text{facet present})$; and (2) a \emph{magnitude regressor}, a ridge regression  (regularization chosen by cross-validation) on a 100-component PCA projection, trained on the pooled GPT-5 labels and the human-labeled nonzero examples to predict an intensity score conditional on presence. Both PCA dimensionalities were selected once on a development split of the Relationships and Accomplishment training data and fixed across facets. Human nonzero examples are weighted relative to GPT-5 examples by a blend weight $w \in \{0,1,10,30,60,100\}$, chosen per facet/outcome via 5-fold cross-validation computed strictly within the training pool, never touching the held-out test set. The final prediction is the product of the two parts, $P(\text{present}) \times \text{magnitude}$. An ablation isolating each component's contribution (Appendix~\ref{sec:full-model-performance}) shows this blending strategy yields substantial gains for Relationship, Accomplishment, and the enduring outcomes of Positive Emotion and Engagement ($\Delta r$ from $+.13$ to $+.50$), but little or no benefit for Meaning and the immediate outcomes of Positive Emotion and Engagement ($\Delta r$ from $-.06$ to $-.13$).

\subsection{Model Validation}
\label{sec:validation}
We first annotated a random sample of 300 reviews. Nonzero labels in this sample are extremely sparse, ranging from 265 (Positive Emotion, Immediate) down to 9 (Accomplishment, Enduring) (Appendix~\ref{sec:nonzero-counts}), too few to evaluate the rarest facets reliably. We therefore construct an enriched test set for each facet: we fit a ridge regression to the GPT-5-annotated reviews, predict scores for a separate pool of 50,000 reviews, and sample 200 equally across the six predicted score bins (0--5; about 33 each), which two annotators then labeled. This yields a more even spread of nonzero labels. At $\alpha=.05$ with 80\% power, the random sample can detect correlations of $r \geq .16$ and each enriched set $r \geq .20$. Performance on the random sample is reported in Appendix~\ref{sec:full-model-performance}.

Table~\ref{tab:model-performance} shows performance on the enriched test sets; qualitative examples of top-scoring reviews per facet are in Appendix~\ref{sec:top-messages}. We assessed inter-annotator reliability on the enriched test sets using the intraclass correlation coefficient (ICC), computing a two-way random effects model with absolute agreement, ICC(2,1), using the \texttt{pingouin} Python package~\citep{vallat2018pingouin}. Because gold labels are the average of two annotators, we also report the Spearman-Brown-adjusted average-rater reliability, ICC(2,2), which we use to deattenuate the correlations in Table~\ref{tab:model-performance} (full values in Appendix~\ref{sec:full-model-performance}, Table~\ref{tab:full-model-performance}). Reliability ranged from ICC(2,1)$=.43$(Meaning, Immediate) to ICC(2,1)$=.75$ (Accomplishment, Immediate). Meaning's comparatively low reliability is consistent with it being the facet hardest for both human annotators and models to agree on throughout this study, despite multiple iterations of the prompt.

Treating the task as zero-vs-nonzero classification, models trained on human annotated data achieve AUCs between .80 and .97 across all facets (Table~\ref{tab:model-performance}), with Meaning again the weakest ($\text{AUC}\approx.80$--$.85$) and Positive Emotion, Immediate the strongest ($\text{AUC}=.965$).

\subsection{Satisfaction Signals and Rankings}
We compare well-being facets against two conventional satisfaction signals: the star rating attached to each review and review sentiment. We score sentiment for all 13.7 million reviews with a pretrained RoBERTa-large sentiment classifier (\textit{siebert/sentiment-roberta-large-english})~\cite{hartmann2023, liu2019roberta}, using the logit of the positive class as each review's sentiment score. To assess how rankings would differ, we correlate per-book means of each signal and rank books (the 16,341 works in our book-content dataset; Section~\ref{sec:datasets}) by these means, computing the share of the top-$K$ books by stars that also appear in the top-$K$ by each other signal ($K \in \{100, 500\}$). Because means for books with few reviews are noisy, we use shrunken means, $\hat{\mu}_b = (n_b\bar{x}_b + m\bar{x})/(n_b + m)$, where $\bar{x}_b$ and $n_b$ are the book's mean score and review count, $\bar{x}$ the corpus mean, and $m=20$ a prior weight, equivalent to adding 20 reviews at the corpus mean.

\subsection{Creating Book Features}
\begin{figure}[t!]
     \centering
     \includegraphics[width=.48\textwidth]{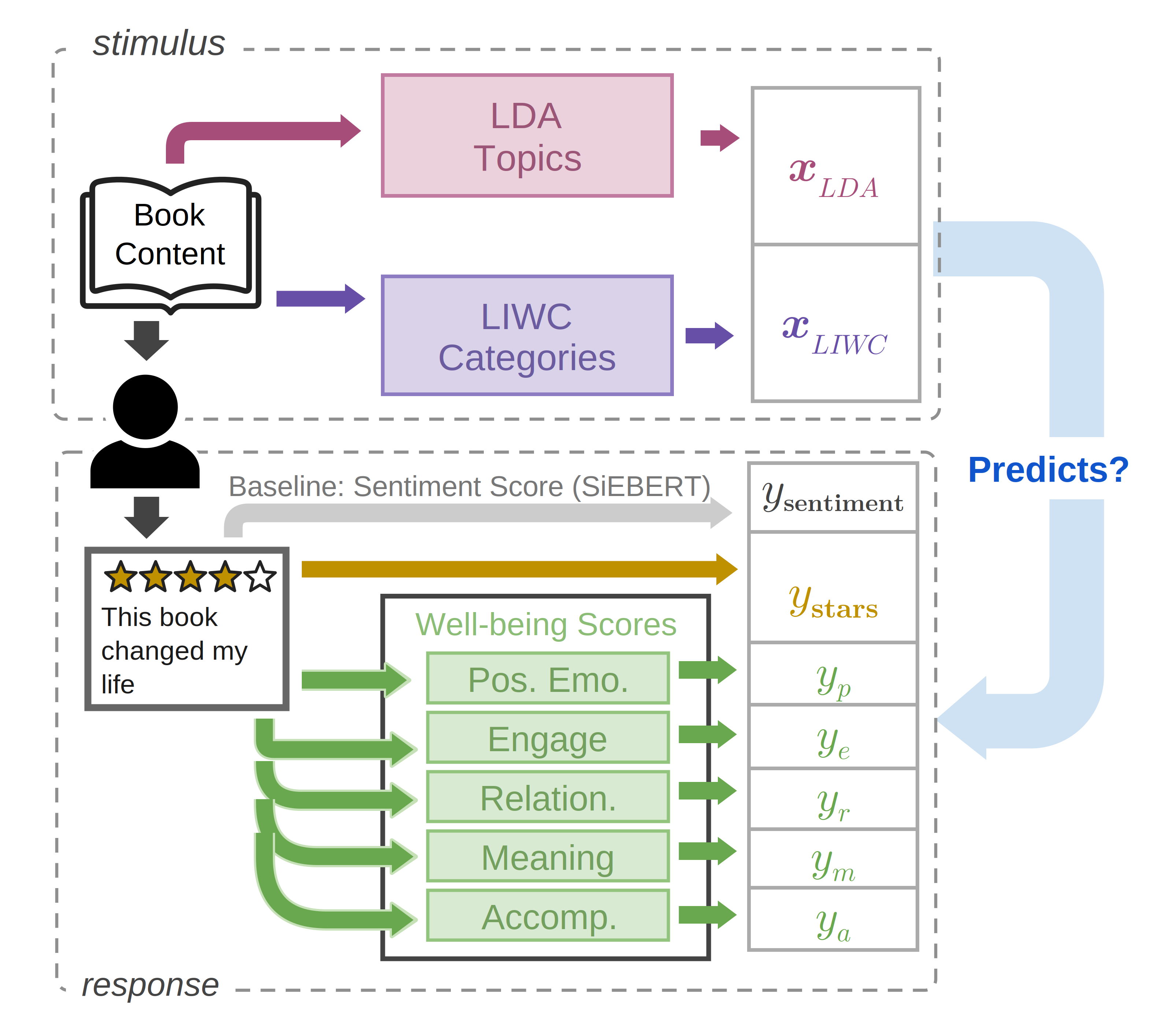}
     \caption{Diagram of RQ2. We correlate stars, sentiment, and PERMA facets with book-level content features. This should allow for analysis of which, and to what degree, stimulus (book content) predicts the response (expressions of well-being by reviewers)}
     \label{fig:ContentFig}
\end{figure}

To understand the book-level content associated with expressions of well-being, we analyze the text of the books themselves. For each work in our book-content dataset (Section~\ref{sec:datasets}), we average review-level star ratings, sentiment, and immediate and enduring facet scores, and use differential language analysis (DLA) to correlate these book-level means with linguistic features extracted from the book text (Figure~\ref{fig:ContentFig}). To characterize book content, we use both open- and closed-vocabulary approaches.

For the open-vocabulary analysis, we project book texts into a 1,000-topic Latent Dirichlet Allocation (LDA) space trained on the entire book corpus ($n=73{,}168$), reduced via non-negative matrix factorization to 300 interpretable supertopics, excluding topics whose top words are $\geq$30\% character names or other proper nouns, which index individual books or series rather than themes. High-dimensional topic models fit on large corpora have been used effectively in prior work \cite{jaidka2020estimating, sear2022dynamic, chen2014topic}. As a closed-vocabulary complement, we compute LIWC category frequencies~\cite{boyd2022development}. We report a subset of categories chosen for face validity with PERMA facets (e.g., Religion, Prosocial, Achievement), along with stylistic categories useful for characterizing writing (pronouns and temporal focus) and Swearing, which showed among the strongest negative associations with well-being. To check whether associations reflect genre composition, we refit selected associations with facet $\times$ genre interaction terms, using each book's top genre tags (Appendix~\ref{sec:genreTables}).

\section{Results}

We first compare well-being facets with conventional satisfaction signals at the review level (RQ1), then examine which features of book content are associated with expressed well-being (RQ2).

\subsection{Satisfaction Signals vs.\ Well-Being}

\begin{table}[htbp]
\small
\centering
\begin{adjustbox}{max width=.5\textwidth}
\setlength{\tabcolsep}{5pt}
\begin{tabular}{llcccc p{3em} p{3em}}
\toprule
\multicolumn{2}{l}{\multirow{2}{*}{\textbf{Language task}}}
  & \multicolumn{2}{c}{\makecell{\textbf{Correlation}\\\textbf{with} \fivestars}}
  & \multicolumn{2}{c}{\makecell{\textbf{Overlap}\\\textbf{with} \fivestars}}
  & \multicolumn{2}{c}{} \\
\cmidrule(lr){3-4} \cmidrule(lr){5-6}
\multicolumn{2}{l}{} &
  \makecell{\textbf{Per}\\\textbf{Review}} & \makecell{\textbf{Avg. Per}\\\textbf{Book}} &
  \makecell{\textbf{Top@100}\\\textbf{Books}} & \makecell{\textbf{Top@500}\\\textbf{Books}} &
  \multicolumn{2}{c}{} \\
\multicolumn{2}{l}{\textit{n}} & 13.7M & 16,341 & 16,341 & 16,341 & \multicolumn{2}{c}{} \\
\midrule

\textbf{Sentiment} & &
  \cellcolor{tealHigh!100!white} .56 & \cellcolor{tealHigh!100!white} .72 &
  \cellcolor{tealHigh!100!white} 38\% & \cellcolor{tealHigh!100!white} 44\% & \multicolumn{2}{c}{}\\

\cline{1-6}
\multirow{2}{*}{\textbf{Pos. Emo.}} & Imm. &
  \cellcolor{tealHigh!86!white} .48 & \cellcolor{tealHigh!48!white} .34 &
  \cellcolor{tealHigh!24!white} 9\% & \cellcolor{tealHigh!44!white} 19\%
  & \multicolumn{2}{c}{\multirow{2}{*}{\makecell{\textit{More} \\ \textit{Hedonic}}}} \\
 & End. &
  \cellcolor{tealHigh!49!white} .28 & \cellcolor{tealHigh!41!white} .30 &
  \cellcolor{tealHigh!16!white} 6\% & \cellcolor{tealHigh!33!white} 14\%
  & \multicolumn{2}{c}{} \\

\cline{1-6}
\multirow{2}{*}{\textbf{Engage.}} & Imm. &
  \cellcolor{tealHigh!80!white} .45 & \cellcolor{tealHigh!39!white} .28 &
  \cellcolor{tealHigh!24!white} 9\% & \cellcolor{tealHigh!45!white} 20\%
  & \multicolumn{2}{c}{} \\
 & End. &
  \cellcolor{tealHigh!38!white} .21 & \cellcolor{tealHigh!28!white} .20 &
  \cellcolor{tealHigh!5!white} 2\% & \cellcolor{tealHigh!22!white} 10\%
  & \multicolumn{2}{c}{} \\

\cline{1-6}
\multirow{2}{*}{\textbf{Relation.}} & Imm. &
  \cellcolor{tealHigh!10!white} .06 & \cellcolor{tealHigh!12!white} .08 &
  \cellcolor{tealHigh!0!white} 0\% & \cellcolor{tealHigh!16!white} 7\%
  & \multicolumn{2}{c}{} \\
 & End. &
  \cellcolor{tealHigh!12!white} .07 & \cellcolor{tealHigh!13!white} .10 &
  \cellcolor{tealHigh!3!white} 1\% & \cellcolor{tealHigh!18!white} 8\%
  & \multicolumn{2}{c}{} \\

\cline{1-6}
\multirow{2}{*}{\textbf{Meaning}} & Imm. &
  \cellcolor{tealHigh!16!white} .09 & \cellcolor{tealHigh!20!white} .15 &
  \cellcolor{tealHigh!18!white} 7\% & \cellcolor{tealHigh!26!white} 12\%
  & \multicolumn{2}{c}{\multirow{2}{*}{\makecell{\textit{More} \\ \textit{Eudaimonic}}}} \\
 & End. &
  \cellcolor{tealHigh!16!white} .09 & \cellcolor{tealHigh!21!white} .15 &
  \cellcolor{tealHigh!13!white} 5\% & \cellcolor{tealHigh!26!white} 11\%
  & \multicolumn{2}{c}{} \\

\cline{1-6}
\multirow{2}{*}{\textbf{Accomp.}} & Imm. &
  \cellcolor{tealHigh!13!white} .07 & \cellcolor{tealHigh!24!white} .17 &
  \cellcolor{tealHigh!11!white} 4\% & \cellcolor{tealHigh!20!white} 9\%
  & \multicolumn{2}{c}{} \\
 & End. &
  \cellcolor{tealHigh!12!white} .07 & \cellcolor{tealHigh!23!white} .17 &
  \cellcolor{tealHigh!8!white} 3\% & \cellcolor{tealHigh!18!white} 8\%
  & \multicolumn{2}{c}{} \\

\bottomrule
\end{tabular}
\end{adjustbox}
\caption{Association of sentiment and PERMA facets with star ratings. \textit{Individual Review}: Pearson $r$ across all $13{,}713{,}881$ reviews. \textit{Avg.\ Per Book}: Pearson $r$ between per-book mean scores. \textit{Top@$K$ Books}: share of the top-$K$ books ranked by each score that also appear in the top-$K$ ranked by stars, using shrunken means to prevent books with few reviews from dominating the rankings (random baseline: 0.6\% at $K{=}100$, 3\% at $K{=}500$). Imm.\ = immediate; End.\ = enduring. Shading is scaled within each column, relative to sentiment. All correlations are significant after Benjamini--Hochberg FDR correction ($q<.05$) \cite{benjamini1995controlling}. Eudaimonic facets are only weakly associated with ratings, and star-based rankings surface few of the books readers associate with eudaimonic well-being.}
\label{tab:rating-correlations}
\end{table}

\begin{table}[htbp]
\footnotesize
\centering
\setlength{\tabcolsep}{3pt}
\renewcommand{\arraystretch}{1.3}
\setlength{\aboverulesep}{0pt}
\setlength{\belowrulesep}{0pt}
\newcommand{\pill}[1]{\tikz[baseline=(t.base)]\node[draw=black!30, rounded corners=2pt, inner xsep=2pt, inner ysep=0.5pt](t){\strut\sffamily #1};}
\begin{tabularx}{\columnwidth}{>{\raggedright\arraybackslash\hangindent=1em\hangafter=1}X}
\toprule
\multicolumn{1}{c}{\textbf{Top 5 Books}} \\
\midrule
\textbf{Stars}\enspace \pill{The Hate U Give} \pill{March: Book Three} \pill{Just Mercy} \pill{Harry Potter \#7} \pill{Harry Potter \#4} \\
\textbf{Sentiment}\enspace \pill{Born a Crime} \pill{Where the Mountain Meets the Moon} \pill{The Best Christmas Pageant Ever} \pill{The Hate U Give} \pill{Just Mercy} \\
\midrule
\rowcolor{rowgray} \textbf{Pos. Emo. Imm.}\enspace \pill{The True Meaning of Smekday} \pill{Can You Keep a Secret?} \pill{Ms.~Marvel, Vol.~1} \pill{Flipped} \pill{Saving Francesca} \\
\rowcolor{rowgray} \textbf{Pos. Emo. End.}\enspace \pill{Peace Is Every Step} \pill{Wherever You Go, There You Are} \pill{Being Peace} \pill{The Power of Now} \pill{The Places That Scare You} \\
\textbf{Engage. End.}\enspace \pill{Jamie's Food Revolution} \pill{Ratio} \pill{The Artist's Way} \pill{Vagabonding} \pill{Betty Crocker's Cookbook} \\
\rowcolor{rowgray} \textbf{Relation. End.}\enspace \pill{Quiet} \pill{All Joy and No Fun} \pill{How Children Succeed} \pill{Mindset} \pill{The Dance of Anger} \\
\textbf{Meaning End.}\enspace \pill{Peace Is Every Step} \pill{Loving What Is} \pill{Living Buddha, Living Christ} \pill{The Artist's Way} \pill{The Power of Now} \\
\rowcolor{rowgray} \textbf{Accomp. End.}\enspace \pill{The First Days of School} \pill{Drawing on the Right Side of the Brain} \pill{Self-Editing for Fiction Writers} \pill{The Non-Designer's Design Book} \pill{Thanks for the Feedback} \\
\bottomrule
\end{tabularx}
\caption{Top 5 books by star rating, sentiment, and PERMA facets, using shrunken per-book means ($n=16{,}341$ books). Immediate lists for Relationships, Meaning, and Accomplishment closely match their enduring lists and are omitted. Satisfaction signals and immediate hedonic facets favor acclaimed and absorbing fiction and memoir, while enduring and eudaimonic facets surface mindfulness, parenting, and skill-building books.}
\label{tab:top-books}
\end{table}

\paragraph{Satisfaction signals align with immediate hedonic well-being, but little else.}
Ratings and sentiment relate most closely to immediate Positive Emotion and Engagement ($r=.48$ and $.45$ with ratings; $.50$ and $.43$ with sentiment), an expected overlap since positive sentiment and positive emotion share language, which offers some validation of our estimates. Even so, these facets share at most a quarter of their variance with satisfaction signals. The immediate/enduring distinction matters only for these hedonic facets: their enduring expressions correlate with ratings about half as strongly ($r=.28$ and $.21$), whereas for Relationships, Meaning, and Accomplishment, immediate and enduring expressions correlate almost identically, and negligibly ($r\leq.09$; e.g., Meaning: $.09$ vs.\ $.09$) (Table~\ref{tab:rating-correlations}). Satisfaction signals thus capture part of momentary hedonic experience, but little of its enduring or eudaimonic dimensions.

\paragraph{Rankings by stars rarely surface high PERMA scoring books.}
To see how this divergence would play out on a web platform, we average reviews per book and rank them. Per-book mean ratings correlate strongly with sentiment ($r=.72$) but much less with any well-being facet ($r\leq.34$). Of the 100 books with the highest average star ratings, 38\% also rank in the top 100 by sentiment, but at most 9\% by any PERMA facet. The top-ranked books also differ in kind (Table~\ref{tab:top-books}): stars and sentiment favor acclaimed fiction and memoir, whereas enduring and eudaimonic facets surface mindfulness, parenting, and skill-building books. A recommender ranking by stars would thus rarely surface the books readers most associate with eudaimonic well-being.

\begin{figure}[t!]
     \centering
     \includegraphics[width=.5\textwidth]{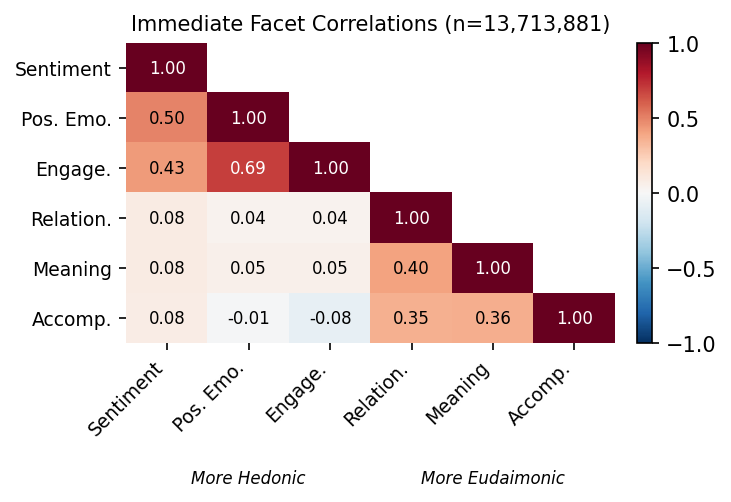}
     \caption{Correlations of Immediate Facet Scores with one another and with Sentiment (full $n=13{,}713{,}881$-review corpus). All pairwise correlations were statistically significant after Benjamini-Hochberg FDR correction (q = .05).  \cite{benjamini1995controlling} See enduring facets in Appendix ~\ref{fig:FacetEndCorrelation}. Hedonic factors associate with each other, eudaimonic factors associate with each other, but hedonic and eudaimonic factors have near zero association.}
     \label{fig:FacetDistributions}
\end{figure}

\paragraph{Facet structure.}
Immediate expressions of well-being cluster along the hedonic/eudaimonic divide (Figure~\ref{fig:FacetDistributions}): immediate Positive Emotion and Engagement correlate strongly ($r=.69$), eudaimonic facets correlate low to moderately with each other ($r=.35--.40$), and hedonic and eudaimonic facets are only weakly related ($r\leq.08$). This separation is clearer than in self-report, where hedonic and eudaimonic well-being overlap substantially~\cite{disabato2016different}, and largely mirrors research on momentary well-being~\cite{heshmati2023momentary}. It is also horizon-specific (e.g. enduring Positive Emotion instead aligns with the eudaimonic facets; Appendix ~\ref{sec:enduringTables}). Engagement is the main departure from prior work, where momentary Engagement was weakly tied to both flourishing and emotional well-being~\cite{heshmati2022assessing}; in book reviews, it aligns closely with Positive Emotion and ratings, plausibly because immediate engagement in a review is overrepresented as absorption in reading, closer to hedonic enjoyment than to PERMA's broader flow construct.

\subsection{Book Content Associated with Well-Being}
We next examine which features of book content are associated with well-being expressed in reviews, using topics (Table~\ref{tab:topics_filtered}) and LIWC categories (Table~\ref{tab:LIWC_Correlations}).

\paragraph{Topics.}
Several facets align with topics that mirror their construct (Table~\ref{tab:topics_filtered}): Relationships with therapy (\emph{patient, client, treatment, self}; $r=.48$--$.50$), Accomplishment with writing craft (\emph{writing, writers, students}; $r=.54$--$.57$), and Meaning with life and the world (\emph{life, world, god, human}; $r=.51$--$.53$). Second-person, possibility-oriented language (\emph{your, can, do, will, if}) is associated positively with enduring expressions of Positive Emotion and every eudaimonic facet ($r=.40$--$.55$).

Immediate Positive Emotion and Engagement stand apart: in 11 of 12 topics, at least one takes the opposite sign from the enduring and eudaimonic facets. Both are flat or negative on every positively associated topic ($r=-.17$ to $.02$) and positive on most inversely associated ones, including narrative action and dialogue (\emph{him, up, out, into, said}; $r$ up to $.16$). Content associated with immediate hedonic expressions thus differs from content associated with enduring and eudaimonic ones. Ratings and sentiment are at most weakly associated with any of the 300 topics ($|r|\leq.16$). The topic most associated with high ratings concerns race and civil rights ($r=.10$), echoing the top-rated books (Table~\ref{tab:top-books}). The topic most inversely correlated with ratings and sentiment is publishing front matter (\emph{isbn, copyright}), likely an edition artifact rather than content.

\begin{table*}[t]
\scriptsize
\centering
\setlength{\tabcolsep}{1.5pt}
\renewcommand{\arraystretch}{1.15}
\newcommand{\tcloud}[1]{\raisebox{-0.5\height}{\includegraphics[width=1.15cm, trim=10 10 10 10, clip]{#1}}}
\begin{adjustbox}{max width=\textwidth}
\begin{tabular}{L{1.0cm} L{0.6cm} *{12}{C{1.2cm}}}
\toprule
\multicolumn{14}{c}{\textbf{Content Topics correlated with Well-being (Pearson $r$)}}\\
\midrule
 & & \multicolumn{6}{c}{\textbf{Positive}} & \multicolumn{6}{c}{\textbf{Inverse}} \\
\cmidrule(lr){3-8}\cmidrule(lr){9-14}
\multicolumn{2}{l}{\textbf{Topic}} &
\tcloud{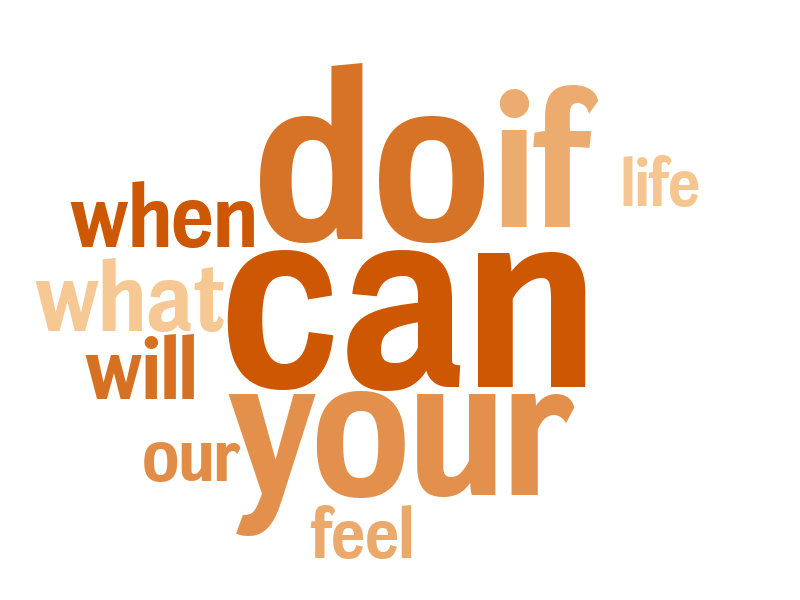} & \tcloud{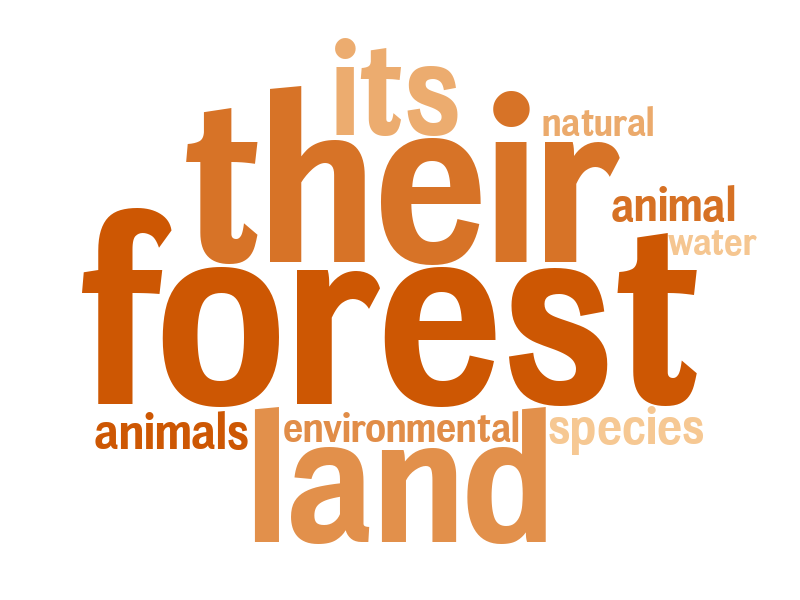} & \tcloud{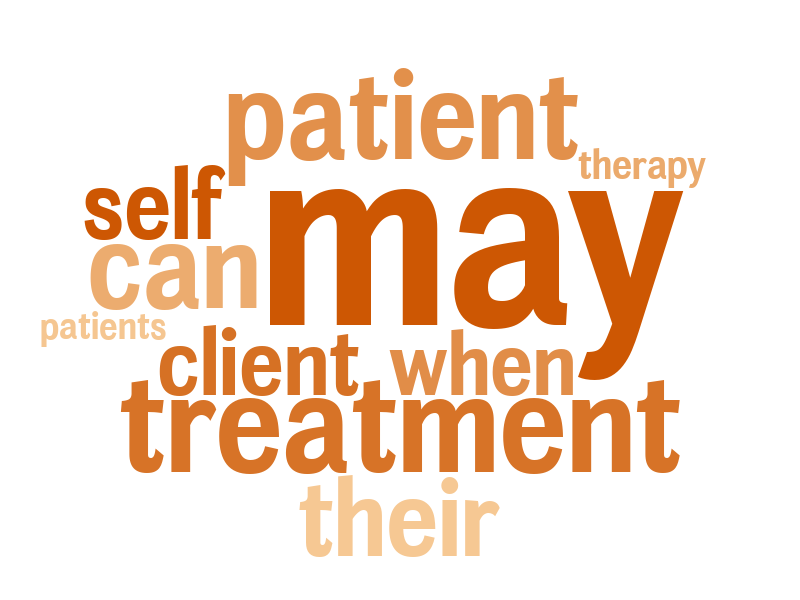} & \tcloud{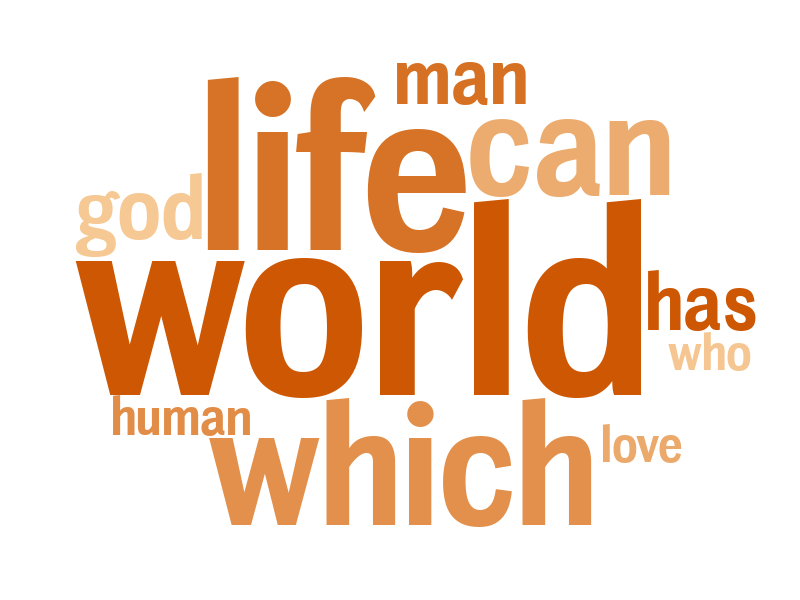} & \tcloud{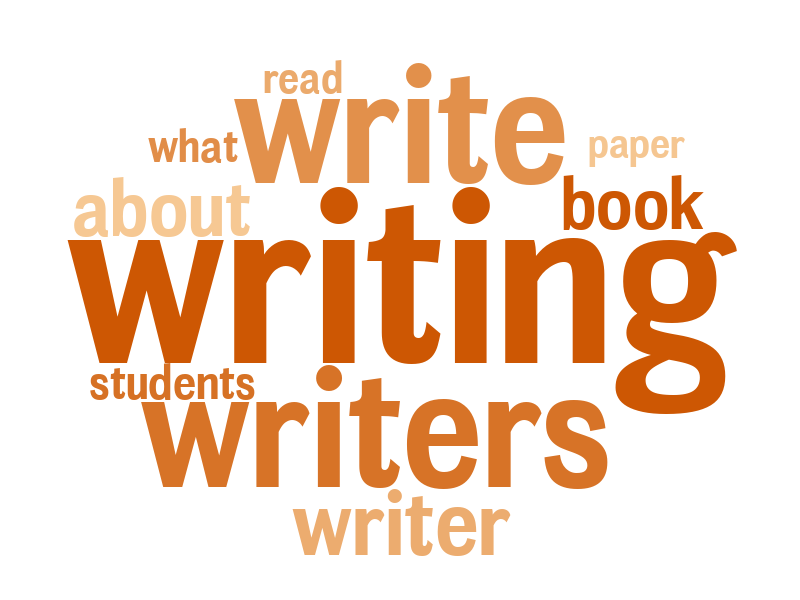} & \tcloud{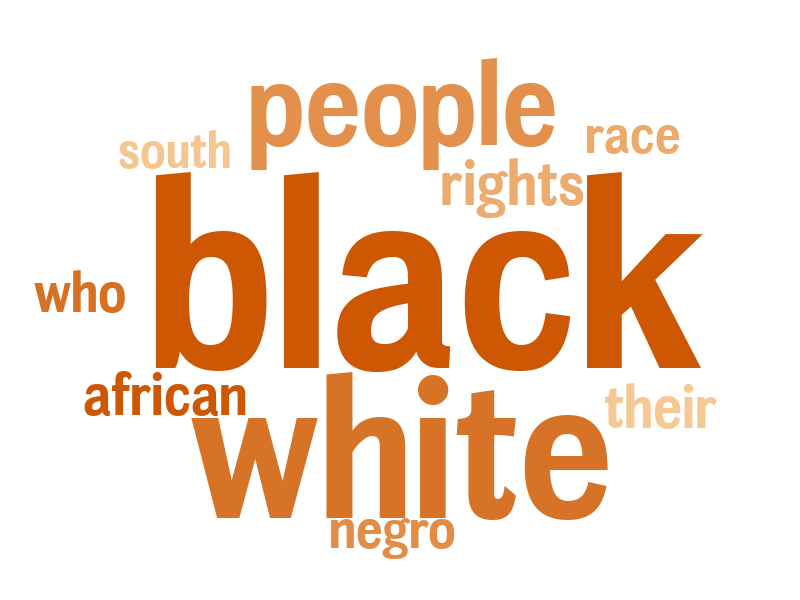} & \tcloud{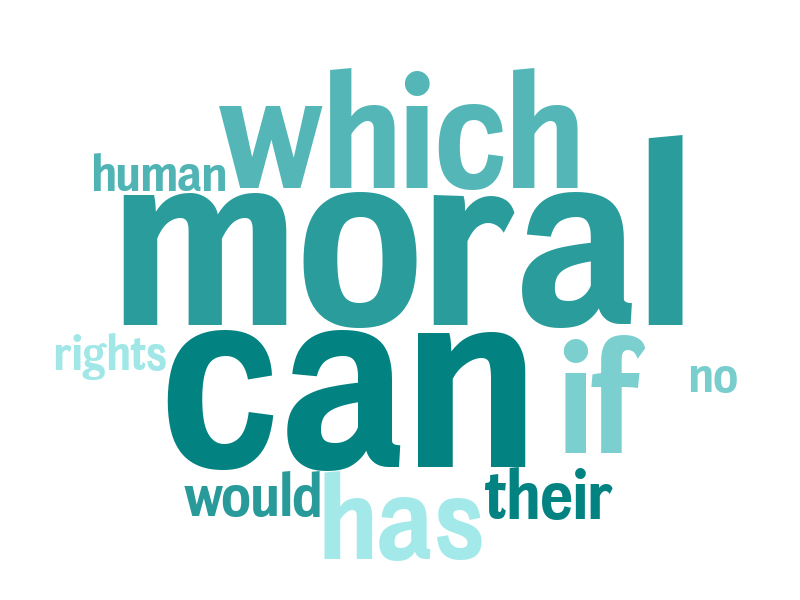} & \tcloud{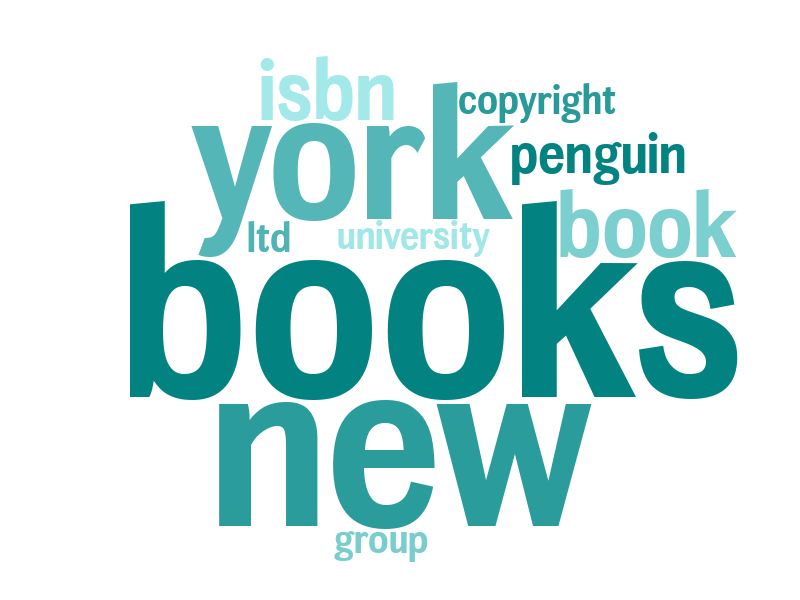} & \tcloud{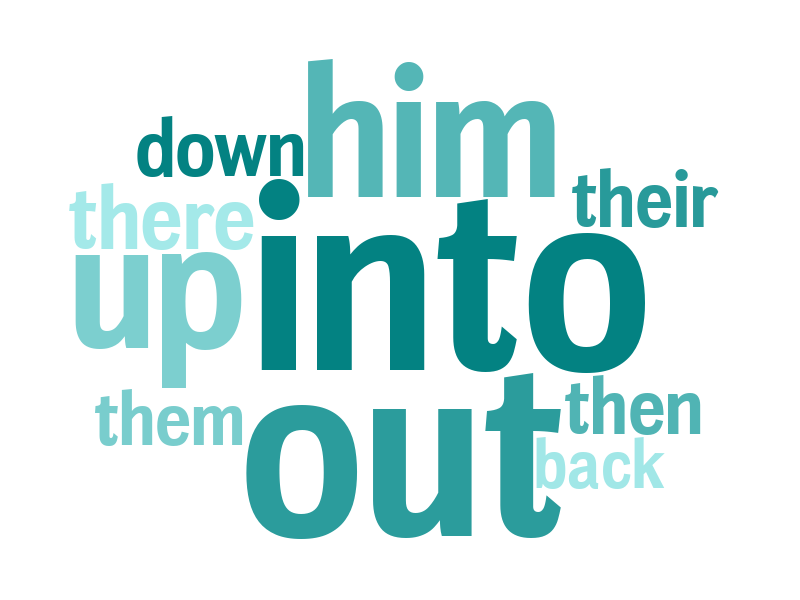} & \tcloud{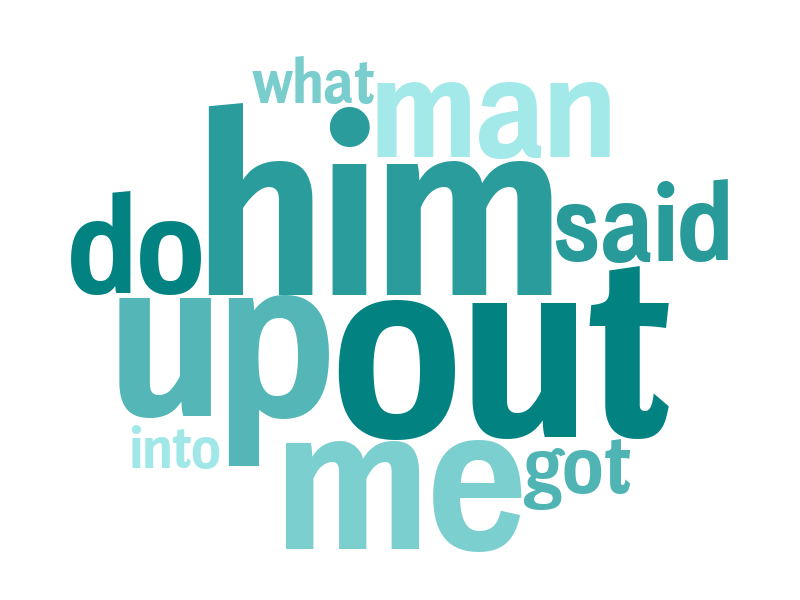} & \tcloud{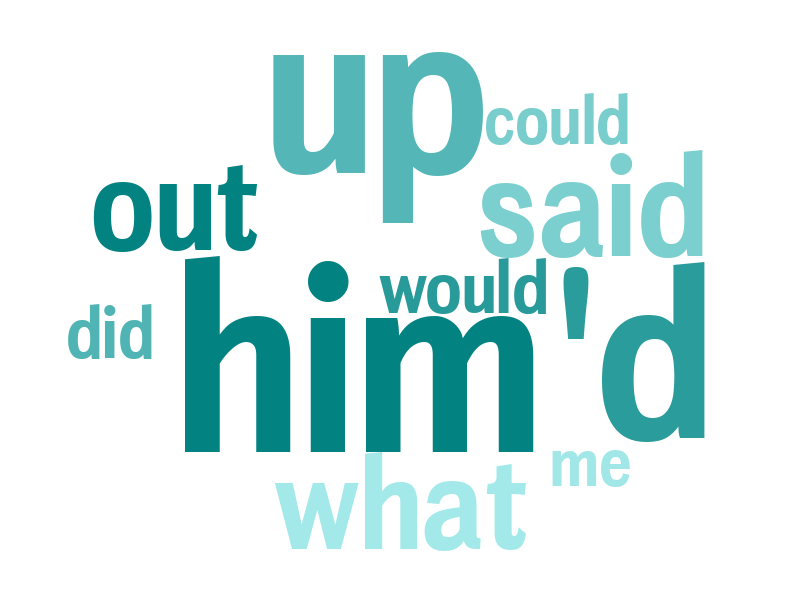} & \tcloud{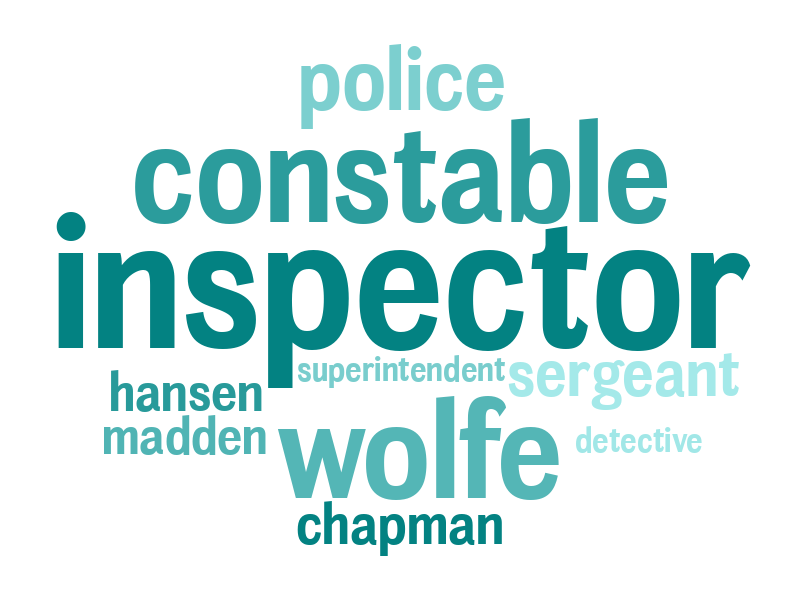} \\
\multicolumn{2}{l}{\textbf{Highest $|r|$ with}} &
\textbf{Pos. Emo.} & \textbf{Engage.} & \textbf{Relation.} & \textbf{Meaning} & \textbf{Accomp.} & \textbf{\fivestars} &
\textbf{Pos. Emo.} & \textbf{Engage.} & \textbf{Relation.} & \textbf{Meaning} & \textbf{Accomp.} & \textbf{\fivestars} \\
\midrule
\textbf{\fivestars} &  & \WCell{\textbf{0.05}}{0.0471} & \WCell{\textbf{0.06}}{0.0647} & \WCell{0.01}{0.0135} & \WCell{\textbf{0.06}}{0.0561} & \WCell{\textbf{0.06}}{0.0568} & \WCell{\textbf{0.10}}{0.1034} & \WCell{\textbf{0.03}}{0.0268} & \WCell{\textbf{0.08}}{0.0791} & \WCell{\textbf{-0.03}}{-0.0276} & \WCell{\textbf{-0.03}}{-0.0305} & \WCell{\textbf{-0.16}}{-0.1573} & \WCell{\textbf{-0.08}}{-0.0815} \\
\textbf{Sentiment} &  & \WCell{\textbf{0.04}}{0.0370} & \WCell{\textbf{0.08}}{0.0793} & \WCell{0.01}{0.0080} & \WCell{\textbf{0.02}}{0.0196} & \WCell{\textbf{0.06}}{0.0611} & \WCell{\textbf{0.10}}{0.0999} & \WCell{0.01}{0.0054} & \WCell{\textbf{0.13}}{0.1312} & \WCell{\textbf{-0.04}}{-0.0408} & \WCell{\textbf{-0.04}}{-0.0380} & \WCell{\textbf{-0.15}}{-0.1507} & \WCell{\textbf{-0.08}}{-0.0817} \\
\midrule
\multirow{2}{*}{\textbf{Pos. Emo.}} & End. & \WCell{\textbf{0.51}}{0.5050} & \WCell{\textbf{0.20}}{0.1965} & \WCell{\textbf{0.27}}{0.2742} & \WCell{\textbf{0.43}}{0.4291} & \WCell{\textbf{0.24}}{0.2367} & \WCell{\textbf{0.11}}{0.1070} & \WCell{\textbf{0.20}}{0.2028} & \WCell{-0.00}{-0.0033} & \WCell{\textbf{-0.09}}{-0.0909} & \WCell{\textbf{-0.15}}{-0.1517} & \WCell{\textbf{-0.11}}{-0.1069} & \WCell{\textbf{-0.18}}{-0.1802} \\
  & Imm. & \WCell{\textbf{-0.04}}{-0.0403} & \WCell{\textbf{0.02}}{0.0241} & \WCell{\textbf{-0.10}}{-0.0963} & \WCell{\textbf{-0.17}}{-0.1675} & \WCell{\textbf{-0.03}}{-0.0289} & \WCell{\textbf{-0.14}}{-0.1351} & \WCell{\textbf{-0.29}}{-0.2896} & \WCell{\textbf{0.25}}{0.2458} & \WCell{\textbf{0.07}}{0.0724} & \WCell{\textbf{0.11}}{0.1119} & \WCell{\textbf{0.13}}{0.1289} & \WCell{\textbf{-0.02}}{-0.0198} \\
\multirow{2}{*}{\textbf{Engage.}} & End. & \WCell{\textbf{0.17}}{0.1708} & \WCell{\textbf{0.29}}{0.2874} & \WCell{\textbf{0.07}}{0.0742} & \WCell{\textbf{0.10}}{0.1000} & \WCell{\textbf{0.23}}{0.2258} & \WCell{\textbf{0.06}}{0.0637} & \WCell{\textbf{0.06}}{0.0598} & \WCell{0.00}{0.0034} & \WCell{\textbf{-0.05}}{-0.0452} & \WCell{\textbf{-0.21}}{-0.2078} & \WCell{\textbf{-0.15}}{-0.1508} & \WCell{\textbf{-0.14}}{-0.1372} \\
  & Imm. & \WCell{\textbf{-0.14}}{-0.1364} & \WCell{\textbf{-0.13}}{-0.1347} & \WCell{\textbf{-0.05}}{-0.0454} & \WCell{\textbf{-0.13}}{-0.1330} & \WCell{\textbf{-0.16}}{-0.1556} & \WCell{\textbf{-0.02}}{-0.0192} & \WCell{\textbf{-0.14}}{-0.1359} & \WCell{\textbf{-0.25}}{-0.2544} & \WCell{\textbf{0.14}}{0.1423} & \WCell{\textbf{0.16}}{0.1621} & \WCell{\textbf{0.25}}{0.2489} & \WCell{\textbf{0.13}}{0.1344} \\
\multirow{2}{*}{\textbf{Relation.}} & End. & \WCell{\textbf{0.42}}{0.4191} & \WCell{\textbf{0.06}}{0.0591} & \WCell{\textbf{0.50}}{0.4965} & \WCell{\textbf{0.11}}{0.1137} & \WCell{\textbf{0.15}}{0.1488} & \WCell{\textbf{0.17}}{0.1675} & \WCell{\textbf{0.17}}{0.1749} & \WCell{\textbf{0.06}}{0.0586} & \WCell{\textbf{-0.15}}{-0.1460} & \WCell{\textbf{-0.09}}{-0.0949} & \WCell{\textbf{-0.07}}{-0.0699} & \WCell{\textbf{-0.12}}{-0.1154} \\
  & Imm. & \WCell{\textbf{0.40}}{0.3988} & \WCell{\textbf{0.06}}{0.0555} & \WCell{\textbf{0.48}}{0.4831} & \WCell{\textbf{0.11}}{0.1122} & \WCell{\textbf{0.14}}{0.1398} & \WCell{\textbf{0.16}}{0.1614} & \WCell{\textbf{0.18}}{0.1791} & \WCell{\textbf{0.05}}{0.0515} & \WCell{\textbf{-0.14}}{-0.1427} & \WCell{\textbf{-0.10}}{-0.0969} & \WCell{\textbf{-0.07}}{-0.0662} & \WCell{\textbf{-0.11}}{-0.1119} \\
\multirow{2}{*}{\textbf{Meaning}} & End. & \WCell{\textbf{0.55}}{0.5535} & \WCell{\textbf{0.07}}{0.0692} & \WCell{\textbf{0.31}}{0.3130} & \WCell{\textbf{0.53}}{0.5347} & \WCell{\textbf{0.24}}{0.2418} & \WCell{\textbf{0.11}}{0.1108} & \WCell{\textbf{0.28}}{0.2758} & \WCell{0.00}{0.0014} & \WCell{\textbf{-0.12}}{-0.1171} & \WCell{\textbf{-0.13}}{-0.1348} & \WCell{\textbf{-0.10}}{-0.0997} & \WCell{\textbf{-0.12}}{-0.1179} \\
  & Imm. & \WCell{\textbf{0.49}}{0.4934} & \WCell{\textbf{0.07}}{0.0750} & \WCell{\textbf{0.28}}{0.2805} & \WCell{\textbf{0.51}}{0.5053} & \WCell{\textbf{0.19}}{0.1913} & \WCell{\textbf{0.12}}{0.1223} & \WCell{\textbf{0.25}}{0.2507} & \WCell{0.01}{0.0099} & \WCell{\textbf{-0.11}}{-0.1094} & \WCell{\textbf{-0.12}}{-0.1206} & \WCell{\textbf{-0.08}}{-0.0784} & \WCell{\textbf{-0.11}}{-0.1142} \\
\multirow{2}{*}{\textbf{Accomp.}} & End. & \WCell{\textbf{0.50}}{0.5020} & \WCell{\textbf{0.06}}{0.0575} & \WCell{\textbf{0.33}}{0.3341} & \WCell{\textbf{0.23}}{0.2317} & \WCell{\textbf{0.54}}{0.5446} & \WCell{\textbf{0.07}}{0.0713} & \WCell{\textbf{0.23}}{0.2311} & \WCell{\textbf{-0.02}}{-0.0196} & \WCell{\textbf{-0.16}}{-0.1611} & \WCell{\textbf{-0.17}}{-0.1668} & \WCell{\textbf{-0.17}}{-0.1711} & \WCell{\textbf{-0.13}}{-0.1279} \\
  & Imm. & \WCell{\textbf{0.53}}{0.5321} & \WCell{\textbf{0.06}}{0.0639} & \WCell{\textbf{0.34}}{0.3428} & \WCell{\textbf{0.26}}{0.2600} & \WCell{\textbf{0.57}}{0.5653} & \WCell{\textbf{0.08}}{0.0850} & \WCell{\textbf{0.23}}{0.2267} & \WCell{-0.01}{-0.0093} & \WCell{\textbf{-0.17}}{-0.1685} & \WCell{\textbf{-0.17}}{-0.1707} & \WCell{\textbf{-0.17}}{-0.1733} & \WCell{\textbf{-0.13}}{-0.1349} \\
\bottomrule
\end{tabular}
\end{adjustbox}
\caption{For each PERMA facet and for Rating, the most positively (left) and most negatively (right) correlated NMF super-topic, selected by the stronger of its immediate (Imm.) or enduring (End.) correlation; each topic shows its correlations with all facets, Rating, and Sentiment ($n=16{,}341$ works). The ``Highest $|r|$ with'' row gives the outcome each topic was selected for. Topics whose top words are $\geq$30\% character names or other proper nouns were excluded. Across all 300 topics, the strongest correlation with ratings is $|r|=.16$ and with sentiment $|r|=.15$. Shading: orange = positive, teal = negative. Bold values are significant after Benjamini-Hochberg FDR correction across all 3{,}600 tests ($q<.05$) \cite{benjamini1995controlling}.}
\label{tab:topics_filtered}
\end{table*}

\paragraph{LIWC categories.}
LIWC categories show similar face validity (Table~\ref{tab:LIWC_Correlations}). Achievement language (\emph{work, better, tried}) has the table's strongest association, with Accomplishment ($r=.46$), and also relates to Relationships and Meaning ($r=.29$--$.34$). Prosocial language (\emph{help, care, sorry}) is among Relationships' strongest associations ($r=.24$--$.25$), and Meaning is associated with first-person plural language ($r\approx.30$) and religion ($r=.25$).

As with topics, writing style separates immediate hedonic expressions from the rest. Third-person singular pronouns (\emph{he, his, she, her}) and past-focused language are negatively associated with every enduring facet ($r$ down to $-.33$ and $-.38$) but positively with immediate Engagement ($r=.39$ and $.37$). To analyze if this was just a function of genre, facet $\times$ genre interaction models were fit. They show that the pronoun associations with Accomplishment and immediate Engagement hold within genre, whereas the association with immediate Positive Emotion reflects genre composition. Swearing follows the same split, negative for every enduring facet ($r=-.10$ to $-.21$) but weakly positive for immediate Positive Emotion and Engagement ($r=.05$ and $.08$). Meaning and Accomplishment instead favor present-focused language ($r=.27$--$.33$); potentially because book tense reflects narrative mode (e.g., present-tense advice versus past-tense story). Ratings and sentiment are again more weakly associated with every displayed category ($|r|\leq.15$).

\begin{table*}[htbp]
\footnotesize
\centering
\setlength{\tabcolsep}{3pt}
\renewcommand{\arraystretch}{1.35}
\setlength{\aboverulesep}{0pt}
\setlength{\belowrulesep}{0pt}
% d & r cell pairs: N = not significant, B = significant (both bold)
\newcommand{\PosN}[3]{\Pos{#1}{#2} & \Pos{#1}{\textcolor{black!50}{#3}}}
\newcommand{\PosB}[3]{\Pos{#1}{\bfseries\boldmath #2} & \Pos{#1}{\bfseries\boldmath\textcolor{black!50}{#3}}}
\newcommand{\NegN}[3]{\Neg{#1}{#2} & \Neg{#1}{\textcolor{black!50}{#3}}}
\newcommand{\NegB}[3]{\Neg{#1}{\bfseries\boldmath #2} & \Neg{#1}{\bfseries\boldmath\textcolor{black!50}{#3}}}
% headers spanning a d/r pair (b = with vertical line after)
\newcommand{\w}[1]{\multicolumn{2}{c}{#1}}
\newcommand{\wb}[1]{\multicolumn{2}{c|}{#1}}
\newcommand{\dr}{\multicolumn{1}{c}{$d$} & \multicolumn{1}{c}{\textcolor{black!50}{$r$}}}
\newcommand{\mn}{\scalebox{0.7}[1]{$-$}}
\newcolumntype{y}{r<{\hspace{-1.5pt}}}
\newcolumntype{z}{>{\hspace{-1.5pt}}r}

\begin{adjustbox}{center,max width=\textwidth}
\begin{tabular}{l l|*{4}{yz}|yz|*{3}{yz}|*{4}{yz}}
\toprule
\multicolumn{26}{c}{\textbf{LIWC Categories of Content correlated with Well-being (Cohen's $d$ / Pearson $r$)}}\\
\midrule
\multicolumn{2}{l|}{\textbf{\textit{Facet / Horizon}}} &
\w{\textbf{Religion}} & \w{\textbf{Prosocial}} & \w{\textbf{Achieve}} & \wb{\textbf{Conflict}} &
\wb{\textbf{Swear}} &
\w{\textbf{Future}} & \w{\textbf{Present}} & \wb{\textbf{Past}} &
\w{\textbf{We}} & \w{\textbf{I}} & \w{\textbf{You}} & \w{\textbf{SheHe}} \\
\midrule
\multicolumn{2}{l|}{\multirow{4}{*}{\textbf{Top Words}}} &
\w{god} & \w{help} & \w{work} & \wb{battle} & \wb{hell} & \w{will} & \w{is} & \wb{was} & \w{we} & \w{i} & \w{you} & \w{he} \\
& & \w{church} & \w{care} & \w{better} & \wb{killed} & \wb{damn} & \w{may} & \w{are} & \wb{had} & \w{our} & \w{my} & \w{your} & \w{his} \\
& & \w{religious} & \w{please} & \w{best} & \wb{attack} & \wb{shit} & \w{might} & \w{can} & \wb{were} & \w{us} & \w{me} & \w{yourself} & \w{she} \\
& & \w{christian} & \w{sorry} & \w{tried} & \wb{enemy} & \wb{bloody} & \w{going} & \w{has} & \wb{been} & \w{ourselves} & \w{myself} & \w{thou} & \w{her} \\
\midrule
& & \dr & \dr & \dr & \dr & \dr & \dr & \dr & \dr & \dr & \dr & \dr & \dr \\
\midrule

\textbf{\fivestars} & &
\PosN{3}{0.04}{.01} & \PosN{1}{0.02}{.01} & \PosB{10}{0.13}{.05} & \PosB{4}{0.07}{.02} &
\NegB{11}{\mn0.14}{\mn.05} &
\NegN{0}{0.00}{\mn.00} & \PosB{6}{0.06}{.03} & \NegB{17}{\mn0.20}{\mn.09} &
\PosB{10}{0.13}{.05} & \NegB{9}{\mn0.12}{\mn.05} & \NegB{9}{\mn0.12}{\mn.05} & \NegB{30}{\mn0.41}{\mn.15} \\
\midrule

\textbf{Sentim.} & &
\PosN{2}{0.03}{.01} & \PosB{4}{0.07}{.02} & \PosB{13}{0.17}{.06} & \NegB{4}{\mn0.06}{\mn.02} &
\NegB{21}{\mn0.30}{\mn.11} &
\NegB{4}{\mn0.05}{\mn.02} & \NegN{1}{\mn0.02}{\mn.00} & \NegB{14}{\mn0.19}{\mn.07} &
\PosB{11}{0.15}{.05} & \NegB{9}{\mn0.11}{\mn.05} & \NegB{17}{\mn0.22}{\mn.08} & \NegB{29}{\mn0.41}{\mn.15} \\
\midrule

\multirow{2}{*}{\makecell[l]{\textbf{Positive}\\\textbf{Emotion}}} & End. &
\PosB{22}{0.19}{.11} & \PosB{30}{0.13}{.15} & \PosB{64}{0.53}{.32} & \NegB{23}{\mn0.41}{\mn.11} &
\NegB{31}{\mn0.46}{\mn.16} &
\PosB{34}{0.26}{.17} & \PosB{57}{0.63}{.28} & \NegB{43}{\mn0.51}{\mn.22} &
\PosB{65}{0.61}{.32} & \PosN{2}{0.07}{.01} & \PosB{10}{\mn0.07}{.05} & \NegB{49}{\mn0.62}{\mn.25} \\
& Imm. &
\NegB{39}{\mn0.48}{\mn.20} & \NegB{6}{\mn0.10}{\mn.03} & \NegB{42}{\mn0.59}{\mn.21} & \NegB{74}{\mn0.98}{\mn.37} &
\PosB{11}{0.17}{.05} &
\NegB{9}{\mn0.13}{\mn.05} & \NegN{2}{\mn0.05}{\mn.01} & \PosB{12}{0.22}{.06} &
\PosB{15}{0.20}{.08} & \PosB{55}{0.78}{.27} & \PosB{46}{0.63}{.23} & \PosB{22}{0.36}{.11} \\
\midrule

\multirow{2}{*}{\textbf{Engage.}} & End. &
\PosB{4}{0.07}{.02} & \NegN{0}{0.07}{\mn.00} & \PosB{43}{0.55}{.21} & \NegB{5}{0.06}{\mn.02} &
\NegB{42}{\mn0.58}{\mn.21} &
\NegB{6}{\mn0.03}{\mn.03} & \PosB{10}{0.10}{.05} & \NegB{41}{\mn0.41}{\mn.20} &
\PosB{6}{0.11}{.03} & \NegB{41}{\mn0.54}{\mn.20} & \NegB{33}{\mn0.44}{\mn.16} & \NegB{63}{\mn0.76}{\mn.31} \\
& Imm. &
\NegN{3}{\mn0.03}{\mn.01} & \PosB{3}{0.03}{.02} & \NegB{32}{\mn0.41}{\mn.16} & \PosB{19}{0.27}{.10} &
\PosB{15}{0.19}{.08} &
\NegB{9}{\mn0.13}{\mn.05} & \NegB{42}{\mn0.58}{\mn.21} & \PosB{74}{1.05}{.37} &
\NegB{7}{\mn0.08}{\mn.03} & \PosB{37}{0.45}{.19} & \PosB{7}{0.06}{.03} & \PosB{78}{1.14}{.39} \\
\midrule

\multirow{2}{*}{\textbf{Relation.}} & End. &
\NegB{9}{\mn0.01}{\mn.05} & \PosB{50}{0.45}{.25} & \PosB{63}{0.69}{.32} & \PosB{4}{\mn0.13}{.02} &
\NegB{20}{\mn0.37}{\mn.10} &
\PosB{15}{0.01}{.08} & \PosB{38}{0.40}{.19} & \NegB{38}{\mn0.45}{\mn.19} &
\PosB{32}{0.45}{.16} & \NegN{1}{0.06}{\mn.01} & \PosN{1}{\mn0.30}{.01} & \NegB{36}{\mn0.59}{\mn.18} \\
& Imm. &
\NegB{7}{0.03}{\mn.04} & \PosB{47}{0.44}{.24} & \PosB{58}{0.66}{.29} & \PosB{6}{\mn0.15}{.03} &
\NegB{19}{\mn0.37}{\mn.09} &
\PosB{14}{0.01}{.07} & \PosB{38}{0.41}{.19} & \NegB{38}{\mn0.45}{\mn.19} &
\PosB{31}{0.44}{.15} & \NegN{1}{0.08}{\mn.01} & \NegN{0}{\mn0.29}{\mn.00} & \NegB{35}{\mn0.57}{\mn.17} \\
\midrule

\multirow{2}{*}{\textbf{Meaning}} & End. &
\PosB{51}{0.41}{.25} & \PosB{42}{0.14}{.21} & \PosB{68}{0.57}{.34} & \NegB{4}{\mn0.32}{\mn.02} &
\NegB{23}{\mn0.44}{\mn.12} &
\PosB{39}{0.25}{.19} & \PosB{60}{0.81}{.30} & \NegB{43}{\mn0.70}{\mn.22} &
\PosB{61}{0.53}{.31} & \NegB{3}{0.09}{\mn.02} & \PosB{16}{\mn0.08}{.08} & \NegB{40}{\mn0.61}{\mn.20} \\
& Imm. &
\PosB{50}{0.38}{.25} & \PosB{39}{0.15}{.19} & \PosB{57}{0.49}{.29} & \NegB{5}{\mn0.35}{\mn.03} &
\NegB{22}{\mn0.38}{\mn.11} &
\PosB{31}{0.18}{.16} & \PosB{54}{0.71}{.27} & \NegB{37}{\mn0.54}{\mn.19} &
\PosB{61}{0.53}{.30} & \PosN{2}{0.17}{.01} & \PosB{11}{\mn0.10}{.06} & \NegB{35}{\mn0.46}{\mn.17} \\
\midrule

\multirow{2}{*}{\textbf{Accomp.}} & End. &
\PosN{1}{0.33}{.00} & \PosB{31}{0.19}{.15} & \PosB{92}{0.96}{.46} & \NegB{4}{0.15}{\mn.02} &
\NegB{25}{\mn0.51}{\mn.12} &
\PosB{41}{0.16}{.20} & \PosB{63}{0.71}{.31} & \NegB{76}{\mn1.32}{\mn.38} &
\PosB{21}{0.20}{.10} & \NegB{35}{\mn0.65}{\mn.17} & \PosB{20}{\mn0.47}{.10} & \NegB{66}{\mn1.58}{\mn.33} \\
& Imm. &
\PosB{4}{0.32}{.02} & \PosB{32}{0.19}{.16} & \PosB{91}{0.90}{.46} & \NegB{7}{0.00}{\mn.03} &
\NegB{25}{\mn0.48}{\mn.12} &
\PosB{41}{0.17}{.21} & \PosB{66}{0.79}{.33} & \NegB{78}{\mn1.34}{\mn.39} &
\PosB{26}{0.28}{.13} & \NegB{31}{\mn0.49}{\mn.16} & \PosB{23}{\mn0.37}{.11} & \NegB{68}{\mn1.56}{\mn.34} \\
\bottomrule
\end{tabular}
\end{adjustbox}
\caption{Columns are LIWC categories; the first four rows in each block provide the category and top example words from that category, computed from each book's own full text with case variants merged (e.g. "He"/"he" pooled) and ranked by corpus-wide frequency. Each category shows Cohen's $d$ (left) (calculated using the top and bottom quartiles of books) and Pearson $r$ (right). The leftmost column shows PERMA facets plus Rating and Sentiment, and the second column indicates enduring (End.) vs immediate (Imm.) where applicable. Outcomes are per-book averages of PERMA scores, star ratings, and sentiment ($n=16{,}341$). Cell shading reflects Pearson $r$ magnitude (global scale $|r|_{\max}=0.50$); teal indicates inverse and orange indicates positive correlation. Bold values indicate that $r$ is significant after Benjamini--Hochberg FDR correction ($q < .05$) \cite{benjamini1995controlling}. Content categories track their matching PERMA facets (e.g., Achieve $\rightarrow$ Accomplishment, Prosocial $\rightarrow$ Relationships, Religion $\rightarrow$ Meaning), swearing is negatively associated with eudaimonic and enduring well-being, and associations shift with tense and person.}
\label{tab:LIWC_Correlations}
\end{table*}

\section{Discussion}
\label{sec:discussion}
Two patterns cut across our analyses. First, enduring Positive Emotion behaves less like its immediate counterpart than like the eudaimonic facets. It correlates strongly with Meaning at the review level ($r=.66$), and in book content it tracks the eudaimonic facets across topics and LIWC categories (e.g., Achieve: $.32$ vs.\ $.34$ for Meaning), while immediate Positive Emotion often takes the opposite sign (Achieve: $-.21$). The hedonic/eudaimonic distinction thus runs within Positive Emotion as well as between facets: momentary enjoyment behaves hedonically, while positive emotion that readers describe as lasting resembles eudaimonic well-being.

Second, book content relates to well-being but barely to satisfaction. Ratings and sentiment show at most weak associations with any content feature ($|r|\leq.16$), whereas well-being facets reach $r=.57$, and the two favor different content. Immediate hedonic expressions, especially Engagement, are associated with narrative action and third-person, past-tense language, while the eudaimonic facets are associated with reflective, present-focused, and goal-directed content such as therapy, writing craft, and reflections on life. Together with the low overlap between star-based and facet-based rankings, this suggests that a system optimizing ratings would favor different content than one optimizing for broader well-being.

\subsection{Implications for Web Platforms}
These findings suggest several ways platforms could move beyond satisfaction signals. First, review language offers a scalable source of richer outcome measures. Platforms already collect reviews at scale, and models like ours could score them for well-being facets as evaluation metrics, revealing what ratings miss, for instance, whether highly rated items are also associated with meaning or accomplishment. Second, facet scores could serve as auxiliary objectives alongside engagement, extending the values-engineering practices platforms already use for individual values~\cite{stray2021you} to a multidimensional, theory-grounded construct. Third, because well-being associations are visible in item content itself, content-based models could estimate well-being-relevant properties of new items before they accumulate reviews, a cold-start setting where satisfaction signals are unavailable.

\subsection{Limitations}
Our analyses are correlational and cannot establish that book content affects reader well-being, and some content associations may partly reflect genre, which we adjust for only in selected analyses. Reviews are a selective window into well-being: reviewers are a self-selected subset of readers, reviews are unprompted, and each captures a single moment, so expressed well-being is a noisy proxy for readers' experience. Identifying well-being in text is also inherently subjective, as reflected in lower inter-annotator agreement for Meaning, and our labels depend on an LLM accessed through an unpinned model alias. Finally, our study covers English-language reviews on one platform in one domain; whether these patterns generalize to other platforms and forms of online content remains open.

\section{Ethical Considerations}
This study analyzes only publicly available, de-identified data released for research use (the UCSD Goodreads dataset); it involved no interaction or intervention with human participants and no collection of identifiable private information, and therefore does not constitute human subjects research. Annotation was performed by members of the research team. Using well-being as an optimization target raises its own concerns: which dimensions to prioritize is a normative choice, frameworks such as PERMA were developed largely in Western contexts and may not reflect all users' values, and well-being signals could be used paternalistically or manipulatively to steer users without their knowledge or consent. We view well-being measures as complements to, not replacements for, user choice, and encourage transparency and user control in any system that adopts them.

\section{Conclusion}
We asked whether the ratings and sentiment that platforms optimize capture eudaimonic, as well as hedonic, well-being expressed in reviews (RQ1), and what book content is associated with each (RQ2). For RQ1, across 13.7 million Goodreads reviews, ratings and sentiment track immediate, hedonic expressions of well-being, relate only weakly to enduring ones, and are nearly unrelated to eudaimonic ones; rankings by stars rarely surface the books readers most associate with eudaimonic well-being.
For RQ2, across 16,341 books, eudaimonic expressions are associated with reflective, present-focused, and goal-directed content, such as therapy, writing craft, and reflections on life, while immediate hedonic expressions follow largely opposite content patterns, and ratings and sentiment register little of either. These results suggest that well-being-aware measures could lead platforms to evaluate and recommend content quite differently than satisfaction signals alone, pointing toward recommender systems that serve not only what users enjoy in the moment but what supports their well-being over time.

\bibliography{custom}

\bibliographystyle{ACM-Reference-Format}

\appendix
\section{Dataset Genre Distribution}
\label{sec:genreTables}

Because broad genres like fiction and non-fiction are common and uninformative on their own, we use each book's top two genre tags. The $n=16{,}341$ review-pooled works are fiction (29.3\%), history/historical fiction/biography (22.1\%), non-fiction (19.9\%), mystery/thriller/crime (6.8\%), fantasy/paranormal (6.6\%), children's (5.6\%), romance (3.0\%), young adult (2.6\%), poetry (2.5\%), and comics/graphic (1.8\%).

\section{Model Details and Full Performance}
\label{sec:full-model-performance}

\begin{table*}[!tp]
\scriptsize
\centering
\providecommand{\hd}[1]{\bfseries\begin{tabular}[b]{@{}c@{}}#1\end{tabular}}
\begin{adjustbox}{max width=\textwidth}
\begin{tabular}{llccccccccc}
\toprule
 & & \multicolumn{2}{c}{\textbf{Reliability}} & & \multicolumn{5}{c}{\textbf{Enriched Test ($N{=}200$ per facet)}} & \hd{Original Test\\($n{=}300$)} \\
\cmidrule(lr){3-4}\cmidrule(lr){6-10}\cmidrule(lr){11-11}
\multicolumn{2}{l}{\textbf{Facet}} &
\hd{ICC\\(2,1)} & \hd{ICC\\(2,2)$^\dagger$} & \hd{$w$} &
\hd{GPT\\Reg.\ $r$} & \hd{GPT Reg.\\+ Clf.\ $r$} & \hd{Full\\$r$} & \hd{Full\\Deatt.\ $r$} & \hd{Full\\AUC} &
\hd{Full\\$r$} \\
\midrule
\multirow{2}{*}{Pos. Emo.} & Imm. & .677 & .807 & 100 & .775 & .743 & .650 & .724 & .965 & .651 \\
 & End. & .574 & .730 & 60 & .549 & .572 & .698 & .817 & .879 & .383 \\
\multirow{2}{*}{Engage.} & Imm. & .570 & .726 & 100 & .665 & .634 & .537 & .630 & .802 & .384 \\
 & End. & .712 & .832 & 30 & .231 & .153 & .493 & .540 & .844 & .300 \\
\multirow{2}{*}{Relation.} & Imm. & .640 & .780 & 1 & .230 & .430 & .734 & .831 & .898 & .420 \\
 & End. & .524 & .687 & 30 & .317 & .393 & .721 & .870 & .925 & .558 \\
\multirow{2}{*}{Meaning} & Imm. & .432 & .603 & 0 & .473 & .464 & .398 & .512 & .849 & .331 \\
 & End. & .589 & .741 & 1 & .421 & .378 & .359 & .417 & .799 & .411 \\
\multirow{2}{*}{Accomp.} & Imm. & .745 & .854 & 100 & .459 & .478 & .700 & .758 & .880 & .510 \\
 & End. & .721 & .838 & 30 & .551 & .489 & .680 & .743 & .883 & n/a \\
\bottomrule
\end{tabular}
\end{adjustbox}
\caption{Model details and full performance. Reliability is inter-annotator agreement on the enriched test set (two-way random effects, absolute agreement); ICC(2,1) is single-rater and ICC(2,2) two-rater-average reliability. $^\dagger$Used to deattenuate the full model's correlations. $w$ is the weight on each human-labeled nonzero example relative to each GPT-5-labeled example, selected by 5-fold CV from $\{0, 1, 10, 30, 60, 100\}$ ($w{=}0$ uses GPT-5 labels only). GPT Reg.\ is the magnitude regressor trained on GPT-5 labels alone; GPT Reg.\ + Clf.\ adds the presence classifier; Full adds the weighted human-labeled examples. Ablation and original-test correlations are uncorrected Pearson $r$; original-test correlations are not deattenuated, as reliability was estimated on the enriched set. The Meaning enriched set has $N{=}189$. Accomplishment-Enduring's original-test $r$ is undefined (see text). Nonzero label rates are in Appendix~\ref{sec:nonzero-counts}.}
\label{tab:full-model-performance}
\end{table*}

Table~\ref{tab:full-model-performance} expands Table~\ref{tab:model-performance} with inter-annotator reliability, the cross-validated blend weights, an ablation of each model component, and performance on an independent, non-enriched sample.

\paragraph{Component ablation.} On the enriched test set, adding the presence classifier to the GPT-5-only regressor improves correlation for 4 of 10 facet-horizons, most notably Relationship-Immediate ($.230 \rightarrow .430$). Adding human-labeled active-learning examples yields the largest gains, improving 6 of 10 facet-horizons, including Engagement-Enduring ($.153 \rightarrow .493$) and both Relationship horizons ($.430 \rightarrow .734$; $.393 \rightarrow .721$). For Positive Emotion-Immediate, Engagement-Immediate, and Meaning, the GPT-5-only regressor achieves higher correlation. We nonetheless apply the same pipeline to every facet. Our central analyses compare facets with one another, and different model types produce differently distributed scores (e.g., the two-part model yields zero-inflated predictions), so selecting models per facet would confound differences between facets with differences between models. 

\paragraph{Score calibration across facets.} Relationship's top predicted scores ($2.50$--$2.58$) are lower than those of the other facets ($3.99$--$4.87$). This reflects calibration rather than weaker signal: Relationship-Immediate's CV-selected weight is $w{=}1$, keeping its regressor close to the conservative GPT-5 labels, whereas Accomplishment, whose GPT-5 label distribution is nearly identical (mean $1.35$ vs.\ $1.34$; about 1.3\% vs.\ 1.2\% of labels $\geq 4$), has $w{=}100$, letting its few high-scoring human examples pull predictions upward. Raw score magnitudes should therefore not be compared across facets.

\paragraph{Original test set.} The original test set is an independent sample of 300 manually annotated reviews drawn by plain random sampling rather than the stratified enrichment in Section~\ref{sec:methods}, with no overlap with the GPT-5 training pool. Its natural distribution is heavily zero-inflated (as few as 9 nonzero Accomplishment-Enduring labels), so its correlations are noisier and generally lower than on the enriched set. Accomplishment-Enduring's $r$ is undefined because the classifier predicted near-zero presence across the entire sample, leaving no variance to correlate.

\section{Enduring Facet Correlations}
\label{sec:enduringTables}

\begin{figure}[t]
     \centering
     \includegraphics[width=\columnwidth]{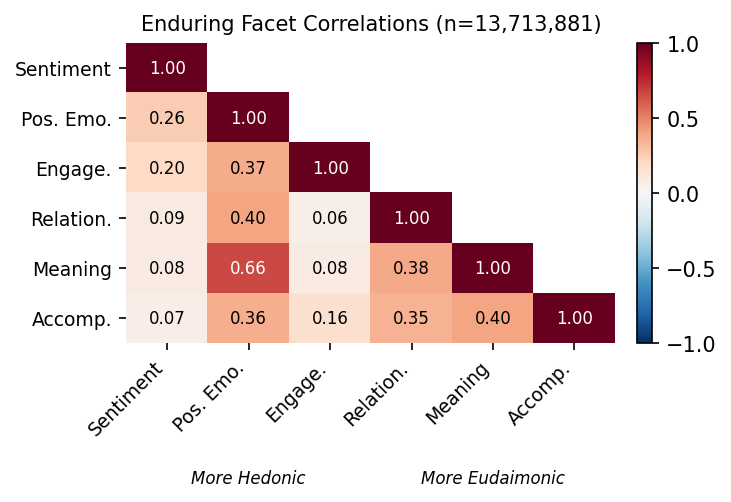}
     \caption{Correlations of enduring facet scores with one another and with sentiment (full $n=13{,}713{,}881$-review corpus). All pairwise correlations are significant after Benjamini-Hochberg FDR correction ($q = .05$; \citealp{benjamini1995controlling}).}
     \label{fig:FacetEndCorrelation}
\end{figure}

Figure~\ref{fig:FacetEndCorrelation} gives the enduring counterpart of the immediate facet correlation figure in the main paper.

\section{Top-Scoring Review Examples by Facet}
\label{sec:top-messages}

Below are the top-2 reviews by predicted immediate score for each facet, drawn from each facet's top-200 scoring reviews corpus-wide and filtered to 100--700 characters. Relationship's lower score range reflects calibration, not weaker signal (Appendix~\ref{sec:full-model-performance}).

{\small
\newcommand{\revcat}[1]{\par\medskip\noindent\textbf{#1}\par\smallskip}
\newcommand{\review}[2]{\par\noindent\hangindent=2em\hangafter=1%
  \makebox[2em][l]{#1}\emph{#2}\par\vspace{3pt}}

\revcat{Positive Emotion}
\review{4.87}{Loved loved loved this book! Super cute and sweet and swoony in all the very best ways. Gideon and Kyle left me with SUCH a huge smile on my face and all the warm, gooey feels. We need lots more of this kind of feel good, low angst LGBTQ book. Super likable characters (well, except one), well written, funny and sweet without being overly sappy. I enjoyed this so much and highly recommend! [URL redacted]}
\review{4.69}{A wonderful, feel-good, laugh out loud book for the Christmas holidays. Don't be misled by the cover, it's more funny rom-com than erotica although there is some heat (but not detail if you know what I mean!) We've all had a bit too much Christmas cheer at times, albeit Peppermintini's in this case and done things we regret. In Samantha's case, it's a dream of revenge on her ex-fiance that goes wrong. This book comes with a HAE and the opening for further books following her two friends. Very well-written and completely enjoyable!}

\revcat{Engagement}
\review{4.56}{This book was just Wow! This book had me hooked from the start and I couldn't wait to see what Cassie would end up doing with er strange ability. The main draw of the story was debating what I would do in her situation. If you knew someone was going to die today would you tell them and try to stop it? What if you did and it caused another tragedy? Would you keep trying? What if your actions to try and prevent their death actually causes their death? Could you live with that? There are no easy answers and it was fascinating seeing the choices Cassie makes and what the people who know about the ability felt she should do.}
\review{4.49}{Not your typical zombies. ARV-3 was amazing. I could not put it down. Not only were the zombies different but the story and society that survived aren't your typical cookie-cutter mold either. I loved Abi and Finn and all of the rest of the characters. I cannot wait to see what happens next! Just one question though Ms. Renae, when will `next' be? I need more!}

\revcat{Relationship}
\review{2.58}{Read this years ago and it changed my life. Recommended for everyone who has or works with kids...or, really, any humans. Great communication advice we could all use and have used on us.}
\review{2.54}{This book works hand in hand with 3 Steps to a Stronger Family. Teaching Your Children Values helps you do just that. It sets up a system for teaching your kids with many suggested activities for different age groups. We've worked our way through this book and have used it to brain storm other ideas. Our children have responded well to the activities and we've all enjoyed the time together. Highly recommended.}

\revcat{Meaning}
\review{4.22}{This book literally changed my life. It started me on my spiritual journey and I'm so grateful for it. I can honestly say that i strive to be the ``awareness'' instead of the mind made little ``me'' every day of my life. Everyone has their share of issues but we are all connected we must love and respect each other,`` one day we will all be corpses and turn into dust ''. This quote is not verbatim but it triggered the much needed shift in my mind. I thankfully haven't been the same since and never will.}
\review{4.16}{EVERYONE NEEDS TO READ THIS! As you can expect, this is a life changing read. It will challenge you, break your heart, and push you to do more and be better for the great good. Every person should read this, especially young Americans. This book challenges you to reconnect to the greater world community. Malala is an inspiration, but she's also a teenage girl. The book reminds me that anyone can be a hero, can do good, and impact others, but also shows that the people who do these things are human too, with struggles, weakness, and silly quirks. If we all had Malala's outlook and drive, this world would be a much better place.}

\revcat{Accomplishment}
\review{4.25}{This book was one of those life-altering books for me. I have a small business and it completely changed my perspective and educated me on what I was doing wrong--and how to fix it. I'm thankful for this book. It gave me clarity and I've since figured out how to re-structure my business so I can remove myself from it, duplicate myself, and plan an exit strategy. I recommend this book for every business owner. I tell people to buy and read it all the time!}
\review{4.18}{I fortunately llanded my first three teaching positions without having to apply myself. That has actually made this job search much more difficult. What did I have to do to prepare myself? Was it diffeent from corpoate America? This book is extremely helpful; everything from resume writing, typical questions and how to work a panel interview. If you're looking for a teaching job, buy this book!}
}

\section{Model Prompt}
\label{sec:model-prompt}
\begin{lstlisting}[style=prompt]
Annotate the immediate and enduring score for each of the following book reviews based on how evident 5 facets, Positive Emotion, Engagement, Positive Relationships, Meaning, or Accomplishment, have each improved for the reviewer as a result of reading the book.

Use the below definitions for each facet.

Positive Emotion (P): Pleasure, rapture, ecstasy, warmth, comfort, happiness, or life satisfaction. It can be in reference to the past (e.g., by cultivating gratitude and forgiveness), the present (e.g., by savoring physical pleasures and mindfulness) or the future (e.g., by building hope and optimism).

Engagement (E): Being absorbed, interested, and involved in an activity or the world itself. Very high engagement is known as "flow": complete absorption in an activity that feels worthwhile in itself, to the point of losing all sense of time. Liking or praising the book alone is Positive Emotion, not Engagement. Enduring Engagement includes lasting engagement with an activity or subject sparked by the book; continuing a series counts only minimally.

Relationships (R): Feeling loved, supported, and valued by others.

Meaning (M): Having a sense of purpose, direction, or value in one's own life, or connecting to something greater than oneself, such as a faith, a charity, or a personally meaningful goal, as reflected in how the reviewer relates the book to their own life. This includes a shift in perspective, values, or priorities; feeling more grateful for one's own life, family, or circumstances; relating the book to one's own faith, purpose, or identity; or feeling moved toward a personally meaningful goal. Enjoying or being moved by a character's story alone is Positive Emotion or Engagement, not Meaning - it must at least be implied that the book affected the reviewer's own sense of meaning.

Accomplishment (A): Objective honors and rewards received, as well as subjective feelings of success, winning, achievement, or mastery. Includes staying on top of daily responsibilities, working towards and reaching goals, and feeling able to complete tasks and daily responsibilities.

Immediate Score (facet_immediate): How much immediate or near-term improvement (within the period of reading or reflecting on the book) is evident for this facet?
0 = No evidence of improvement due to reading the book
1-2 = Minor or subtle near-term improvement due to reading the book
3 = Clear, moderate improvement soon after reading due to reading the book
4-5 = Strong or transformative immediate improvement due to reading the book

Enduring Improvement (facet_enduring): How much deep, long-term, or life-changing improvement in this facet (a fundamental shift in worldview, habits, or identity) is there for the reviewer as a result of reading the book?
0 = No evidence of fundamental change due to reading the book
1-2 = Slight or partial change in perspective or values due to reading the book
3 = Noticeable, meaningful long-term improvement due to reading the book
4-5 = Profound or transformative life improvement due to reading the book

Confidence Scale (for intensity):
0-20 = Very Low (speculative; little or no evidence)
21-40 = Low (weak or vague evidence; uncertain)
41-60 = Moderate (some support, but ambiguous)
61-80 = High (clear, direct evidence; little ambiguity)
81-100 = Very High (explicit, detailed, and unambiguous evidence of change)

Respond only in the format below for each of the 5 facets:

<An integer immediate facet score from 0-5>
<An integer enduring facet score from 0-5>
<An integer confidence of your score, 0-100>

Output format (facet_immediate, facet_enduring, confidence) for 5 facets in order:
Return the result by calling the function 'record_scores' with exactly these fields:
P_immediate, P_enduring, P_conf,
E_immediate, E_enduring, E_conf,
R_immediate, R_enduring, R_conf,
M_immediate, M_enduring, M_conf,
A_immediate, A_enduring, A_conf.
Do not include explanations.

REVIEW:
\end{lstlisting}

\section{Nonzero Label Rates}
\label{sec:nonzero-counts}
\begin{table}[H]
\small
\centering
\begin{tabular}{lcccc}
\toprule
 & \multicolumn{2}{c}{\textbf{Enriched}} & \multicolumn{2}{c}{\textbf{Random Sample}} \\
\textbf{Facet} & \textbf{Imm.} & \textbf{End.} & \textbf{Imm.} & \textbf{End.} \\
\midrule
Positive Emotion & 74.0\% & 49.5\% & 88.0\% & 52.2\% \\
Engagement        & 76.5\% & 62.5\% & 39.2\% & 86.4\% \\
Relationships     & 31.5\% & 22.5\% & 11.3\% &  9.0\% \\
Meaning           & 69.3\% & 47.1\% & 33.2\% & 27.9\% \\
Accomplishment    & 41.0\% & 36.0\% &  3.7\% &  3.0\% \\
\bottomrule
\end{tabular}
\caption{Percentage of reviews with a nonzero human label, in the enriched test set ($n=200$) and in \texttt{FinalTestReviews} ($n=300$, plain random sample).}
\end{table}

\end{document}